\documentclass{aastex}
\usepackage{rotating}

\shorttitle{Herschel Observation of Submillimeter Galaxies at $z>$ 4 }
\shortauthors{Huang et al.}

\begin{document}


\title{HerMES: Spectral Energy Distributions of Submillimeter Galaxies at $z>$4\thanks{Herschel is an ESA space observatory with science instruments provided by European-led Principal Investigator consortia and with important participation from NASA}}

\author{J.-S. Huang\altaffilmark{1,2,3}, D. Rigopoulou \altaffilmark{4,5}, G. Magdis\altaffilmark{4}, M. Rowan-Robinson\altaffilmark{6}, Y. Dai\altaffilmark{3,7}, J.J. Bock\altaffilmark{8}, D. Burgarella\altaffilmark{9},
S. Chapman\altaffilmark{10}, 
D.L. Clements\altaffilmark{6}, A. Cooray\altaffilmark{11}, D. Farrah\altaffilmark{12},
J. Glenn \altaffilmark{13}, S. Oliver\altaffilmark{14}, A. J. Smith\altaffilmark{14}, L. Wang\altaffilmark{14}, M. Page\altaffilmark{15}, D. Riechers\altaffilmark{16,17},
I. Roseboom\altaffilmark{18},
M. Symeonidis\altaffilmark{15}, G. G. Fazio\altaffilmark{3}, M. Yun\altaffilmark{19}, T. M. A. Webb\altaffilmark{19}, A. Efstathiou\altaffilmark{21}} 
\altaffiltext{1}{National Astronomical Observatories of China, Chinese Academy of Sciences, Beijing 100012, China}
\altaffiltext{2}{China-Chile Joint Center for Astronomy, Chinese Academy of Sciences, Camino El Observatorio, \#1515,
Las Condes, Santiago, Chile}
\altaffiltext{3}{Harvard-Smithsonian Center for Astrophysics, 60 Garden Str.,  Cambridge, MA02138, USA}
\altaffiltext{4}{Department of Physics, Denys Wilkinson Building, Keble Road, Oxford, OX1 3RH, UK}
\altaffiltext{5}{RAL Space, Science \& Technology Facilities Council, Rutherford Appleton Laboratory, Didcot, OX11 0QX, UK}
\altaffiltext{6}{Astrophysics Group, Imperial College London, Blackett Laboratory, Prince Consort Road, London SW7 2AZ, UK}
\altaffiltext{7}{Boston College, Newtown, MA, USA}
\altaffiltext{8}{California Institute of Technology, 1200 E. California Blvd., Pasadena, CA 91125, USA; Jet Propulsion Laboratory, 4800 Oak Grove Drive, Pasadena, CA 91109, USA}
\altaffiltext{9}{Laboratoire d'Astrophysique de Marseille, OAMP, Universit� Aix-marseille, CNRS, 38 rue Fr�d�ric Joliot-Curie, 13388 Marseille Cedex 13, France}
\altaffiltext{10}{Institute of Astronomy, University of Cambridge, Madingley Road, Cambridge CB3 0HA}
\altaffiltext{11}{Department of Physics and Astronomy, University of California, Irvine, CA 92697, USA}
\altaffiltext{12}{Department of Physics, Virginia Tech, Blacksburg, VA, 24061, USA}
\altaffiltext{13}{Department of Astrophysical and Planetary Sciences, CASA 389-UCB, University of Colorado, Boulder, CO 80309, USA}
\altaffiltext{14}{Astronomy Center, Department of Physics and Astronomy, University of Sussex, Falmer, Brighton BN1 9QH, UK}
\altaffiltext{15}{Mullard Space Science Laboratory, University College London, Holmbury St Mary, Dorking, Surrey RH5 6NT, UK}
\altaffiltext{16}{Astronomy Department, California Institute of Technology, MC 249-17, 1200 East California Boulevard, Pasadena, CA 91125, USA}
\altaffiltext{17}{Department of Astronomy, Cornell University, 220 Space Sciences Building, Ithaca, NY 14853, USA}
\altaffiltext{18}{Institute for Astronomy, University of Edinburgh, Royal Observatory, Blackford Hill Edinburgh, EH9 3HJ}
\altaffiltext{19}{Department of Astronomy, University of Massachusetts, 710 North Pleasant Street, Amherst, MA 01003, USA}
\altaffiltext{20}{Department of Physics, McGill University, Rutherford Physics Building, 3600 rue University, Montreal, QC H3A 2T8, Canada}
\altaffiltext{21}{School of Sciences, European University Cyprus, Diogenes Street, Engomi, 1516 Nicosia, Cyprus.}



\begin{abstract}

  We present a study of the infrared properties for a sample of seven spectroscopically confirmed submillimeter galaxies at $z>$4.0. By combining ground-based near-infrared, Spitzer IRAC and MIPS, Herschel SPIRE, and ground-based submillimeter/millimeter photometry, we construct their Spectral Energy Distributions (SED) and a composite model to fit the SEDs. 
The model includes a stellar emission component at $\lambda_{\rm rest} <$ 3.5$\,\mu$m; a hot dust component peaking at $\lambda_{rest} \sim$ 5$\,\mu$m; and
cold dust component which becomes significant for $\lambda_{\rm rest} >$ 50$\,\mu$m.  Six objects in the sample are detected at 250 and
350$\,\mu$m.  The dust temperatures for the sources in this sample are
in the range of 40$-$80 K, and their $L_{\rm FIR}$ $\sim$ 10$^{13}$ L$_{\odot}$ qualifies them as Hyper$-$Luminous Infrared Galaxies (HyperLIRGs).  The mean FIR-radio index for this sample is around $< q > = 2.2$  indicating no radio excess in their radio emission. Most sources in the sample have 24$\,\mu$m detections corresponding to a rest-frame 4.5$\,\mu$m luminosity
of Log$_{10}$(L$_{4.5}$/L$_{\odot}$) = 11 $\sim$ 11.5. Their L$_{\rm 4.5}$/$L_{\rm FIR}$ ratios are very similar to those of starburst dominated submillimeter galaxies at $z \sim$ 2. 
The $L_{\rm CO}-L_{\rm FIR}$ relation for this sample is consistent with that determined for local ULIRGs and SMGs at $z \sim$ 2.  We conclude that submillimeter galaxies at $z >$ 4
are hotter and more luminous in the FIR, but otherwise very similar to those at $z \sim$ 2. 
None of these sources show any sign of the strong QSO phase being triggered.
\end{abstract}


\section{Introduction}

Submillimeter observations provide a very effective way of detecting luminous galaxies at high redshifts. The well-known
negative K$-$correction for galaxies in the FIR and submillimeter bands compensates the distance effect, making submillimeter observations
equally sensitive to infrared luminous galaxies in a very wide redshift range of 2 $< z <$ 10 \citep{blain2002}.  
Most SubMillimeter Galaxies (SMGs) are known to be Ultra-Luminous InfraRed Galaxies (ULIRGs, 10$^{12}$ L$_{\odot}$ $<$ $L_{\rm IR}$ $<$ 10$^{13}$ L$_{\odot}$) or Hyper-luminous
Infrared Galaxies (HyperLIRGs, $L_{\rm IR} >$ 10$^{13}$ L$_{\odot}$), implying very intense star forming activity with SFR $>$ 100 M$_{\odot}$/yr. 
Theoretical studies \citep{chakrabarti2008, narayanan2010}
suggested that galaxy-galaxy major mergers can produce  such an intensive star forming activity.
The major-merger scenario for ULIRGs predicts that a massive black-hole is also formed during merging.
A ULIRG will eventually turn into a classical QSO after feedback from the central massive black hole repels most of its gas in the system and quenches star formation. A large spectroscopic survey for SMGs with 
radio detection obtained the first SMG redshift sample at $z \sim 2.2$ (e.g. Chapman et al. 2003). Reproducing number densities and FIR flux densities for this population has been a major challenge in the current theoretical $\Lambda$CDM frame \citep{baugh2005,swinbank2008}. \citet{swinbank2008} proposed to adopt a flat Initial Mass Function (IMF) for SMGs
to match their redshift distribution and number counts.

Understanding the formation of higher redshift SMGs ($z > 4$) is even more challenging.
High-redshift SMGs appear to be more luminous, for example, SMG GN20, the brightest source at 
850$\,\mu$m in the GOODS North field, has a spectroscopic redshift of $z = 4.05$. 
All detected SMGs at $z > 4$ have millimeter flux densities of 5-10$\,m$Jy \citep{younger2007}. At this redshift the millimeter band samples the same 
rest-frame wavelength as the submillimeter band for SMGs at $z\sim$2. These SMGs would then have $f_{\rm 850}$ = 10 $\sim$ 20mJy if they were at $z\sim$ 2. The probability to find such a halo at $z >$ 4 is much lower that at $z \sim$ 2.
Yet more and more high-redshift SMGs have been identified  through multi-wavelength photometry, permitting studies of their formation mechanism through their SEDs and other observed properties.

 Observational identification of SMG counterparts at high redshifts has been very difficult. A definite breakthrough was made with submillimeter and millimeter interferometric observations.
\citet{iono2006}, \citet{younger2007} and \citet{wang2007} used the SubMillimeter Array (SMA) to detect radio-quiet millimeter sources in both GOODS-N and COSMOS fields with a 2\arcsec\ angular resolution. The high resolution submillimeter observations of these sources permit identification of their optical and infrared counterparts. These sources are found to be B$-$band dropouts and have much fainter IRAC flux densities than $z \sim$ 2 SMGs in \citet{chapman2003}, in agreement with these source being at $z >$ 4. \citet{wang2009} performed ultra-deep HST H$-$band observations of GN10, an SMG in the GOODS-North region,  
and yet did not detect any H-band counterpart. With such an extremely red color of H $-$ [3.6] $>$ 4.0,
\citet{wang2009} argued that GN10 must be at $z >$ 4. 

  The SMGs in this study are all spectroscopically identified in 4$< z <$ 5.3.
The first high redshift SMG, GN20, was identified at $z = 4.05$ through observations of the CO[4$-$3] transition using the Plateau de Bure interferometer \citep{daddi2009b}.  A nearby radio source, GN20.2, was subsequently identified at $z=4.05$ through detection of the same CO transition. 
Finally, the extremely red SMG, GN10, was also found to be at $z = 4.04$ through the CO[4$-$3] line \citep{daddi2009a}. 
There was also substantial effort in performing deep optical spectroscopy for optically faint millimeter sources in the COSMOS field 
using the Keck telescopes by \citet{capak2008, capak2011, smolcic2011}, and obtained optical spectroscopic redshifts 
for Capak4.55, AzTEC1 and AzTEC3 at $z = 4.55$, 4.64 and 5.3, respectively.
Subsequent millimeter spectroscopic observations detect of the CO[4$-$3] line from Capak4.55 \citep{schinnerer2008} and CO[2$-$1], CO[5$-$4], 
and CO[6$-$5] lines from AzTEC3 \citep{riechers2010}, confirming their optical spectroscopic redshifts. There was no CO detection
from AzTEC1 in the band inferred from its optical spectroscopic redshift \citep{smolcic2011}. AzTEC1
shows only UV absorption lines in its optical spectrum.  Additionally, AzTEC1 is a B$-$band dropout and has a very low radio-to-submillimeter flux density  ratio, both
of which are consistent with the source being at $z >$ 4 \citep{younger2007}. Finally, another SMG, LESS J033229.4-275619, in the  GOODS-S field was identified
at $z = 4.75$ with both optical spectroscopy and millimeter CO[2-1] line  followup observations\citep{coppin2009}.

Before the advent of the Herschel Space Observatory (hereafter Herschel), our knowledge of the shape of the FIR SED from these galaxies was only based on the available submillimeter and millimeter bands ($\lambda > 850 \mu$m). 
Additional FIR photometry at shorter wavelengths is required together 
with submillimeter and millimeter photometry for the measurement of
important physical parameters, such as dust temperature, dust mass and FIR luminosities. It is, however, extremely difficult to
perform 350 or 450$\,\mu$m observations from ground-based submillimeter telescopes. \citet{kovacs2006} carried out 350$\,\mu$m
observations for a small sample of SMGs at 1 $< z <$ 3 and obtained a mean dust temperature of 34 K  and a mean radio-to-FIR ratio of $q =2.14$.
They concluded that SMGs in their sample are star-burst dominated ULIRGs with no significant AGN contributions.

Herschel carries out effective observations in the FIR bands up to 500$\,\mu$m. The Spectral and Photometric Imaging REceiver\citep[SPIRE]{Griffin2010}
on board Herschel \citep{pilbratt2010} can perform imaging and spectroscopy in 250-500$\,\mu$m bands, which probes the peak of galaxy FIR SED at $z =1$ $\sim$ 3. This is the first time that we are able to carry out surveys at 250, 350 and 500$\,\mu$m. A large number of FIR selected galaxies have been obtained through the Herschel Multi-tiered Extragalactic Survey (HerMES, Oliver et al. 2012, 2010\footnote{hermes.sussex.ac.uk}).  \citet{magdis2010} showed that SPIRE
can easily detect ULIRGs at $z \sim$ 2. At these redshifts galaxies with faint MIPS 24$\,\mu$m emission are subject to the effects of confusion and require  very careful work in extracting photometry from the Herschel SPIRE images \citep{rigopoulou2010}. \citet{magnelli2012} performed a FIR study for a large sample of 61 SMGs in a
wide redshift range. These studies confirm that Herschel SPIRE is very sensitive in detecting  galaxies at higher redshifts. 

In this paper, we present a multi-wavelength study of SEDs for a sample of SMGs at $z > 4$ in the HerMES fields with available full optical and IR SEDs and high resolution
interferometric measurements which are used to refine the photometry and break the confusion. In \S 2, we describe the procedure used
to measure flux densities for the $z >$ 4 sources from the Herschel 250, 350, and 
500$\,\mu$m images for the SMG sample. We present SEDs and derive FIR physical parameters for the sample in \S3 and conclusions in \S 4. We adopt a standard cosmological geometry of
h = 0.7, $\Omega_M$ = 0.3, and $\Omega_{\Lambda}$ = 0.7 throughout this paper.

\section{Spitzer and Herschel Observations of High Redshift SMGs}
\subsection{Spitzer Deep Imaging of High Redshift SMGs}

Most studies of high redshift SMGs \citep{younger2007} relied on their mid- infrared (mid-IR) SEDs to determine their properties and photometric redshifts.
Several groups \citep{huang2004, farrah2008, huang2009, desai2009} successfully used the 4 IRAC bands to identify ULIRGs, so-called "bumpers",  
at $z \sim$ 2 based on the 1.6$\,\mu$m bump shifting in either the 4.5 or 5.8$\,\mu$m band. At $z >$ 4, the 1.6$\,\mu$m bump shifts 
beyond the 8$\,\mu$m band.  Similarly, we can use 5.8, 8.0, and 16$\,\mu$m photometry to classify SEDs for  SMGs at $z >$ 4.
 
The present sample consists of 7 SMGs at $z >$ 4 located in the GOODS-N, GOODS-S, and COSMOS fields.
There are ultra-deep IRAC, IRS-peakup 16$\,\mu$m, and MIPS 24$\,\mu$m images in the GOODS-N and GOODS-S fields.
In addition to the existing deep IRAC imaging in both GOODS fields, the Spitzer Extended Deep Survey (SEDS) carried out in the Spitzer post-cryo mission, covers the GOODS fields at 3.6 and 4.5$\,\mu$m with exposure time of 12 hours per pointing. \citet{ashby2012} combined all IRAC data available in the GOODS-N field 
including GOODS, SEDS, Spitzer GTO0008, and GO20218 programs to achieve better sensitivity and more accurate photometry.
\citet{teplitz2006} performed a deep 16$\,\mu$m imaging in the GOODS-N and GOODS-S field reaching down to a 3$\sigma$ limiting flux
density of 50$\,\mu$Jy. The MIPS 24$\,\mu$m image for the GOODS-N includes data from both GOODS and FIDEL with a total exposure time per position of 10 hours and reaching down to a 3$\sigma$ limiting flux density of 30$\,\mu$Jy.
The Spitzer-COSMOS project \citep[SCOSMOS] {sanders2007} includes both IRAC
and MIPS images \citep{sanders2007}, but these are too shallow to detect the IR counterparts of the sources in this sample.  
Subsequently, we were awarded a total of 37.4 hours with a Spitzer GO program (ID:40801) to carry out deep mid-IR imaging for 4 sources in the COSMOS field. The exposure times for each object are  4.5 hours in the 4 IRAC bands, 3.2 hours in the IRS peakup 16$\,\mu$m band, and 3.8 hours in the MIPS 24$\,\mu$m band.  

All objects in our sample have been detected in submillimeter or radio interferometric observations: five  were detected
by SMA \citep{iono2006,younger2007,wang2009}; while the remaining two objects are also detected at 1.4 GHz. Their
accurate positions in submillimeter or radio bands (Table 1) permit identification of their counterparts in shorter 
wavelength bands. All objects are detected at IRAC 3.6$-$8$\,\mu$m and MIPS 24$\,\mu$m bands. Although, 16$\,\mu$m observations are critical in differentiating between AGN and starburst SEDs for galaxies at $z >$4,  only one object, AzTEC1, has a firm 16$\,\mu$m detection. 
The 16$\,\mu$m limiting flux densities were used to constrain SEDs for the remaining objects.
Finally, deep NIR images are also available in the COSMOS and GOODS-N fields \citep{mccracken2010, wang2009}. All objects in our sample are detected in the NIR bands
except GN10, which is not detected in the deepest available H-band image in the GOODS-N field \citep{wang2009}. The near-IR and mid-IR photometry data for all objects in the sample are shown
in Table 2.

\subsection{Herschel SPIRE Photometry of High Redshift SMGs}

\begin{figure}
\includegraphics[height=160mm, width=120mm,angle=-90]{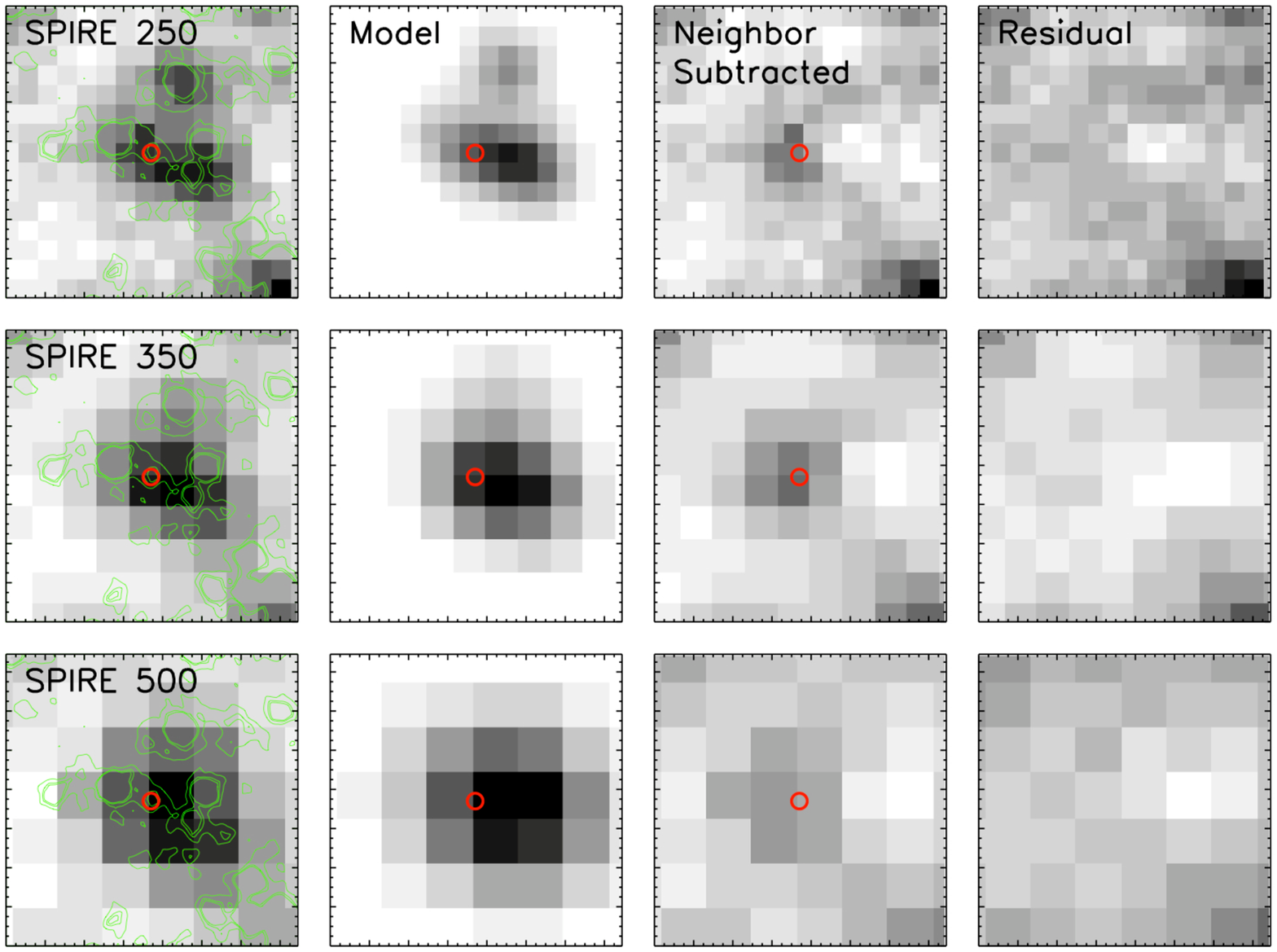}
\caption{The SPIRE stamp images of GN20 for source extraction and photometry. The SPIRE 250, 350 and 500$\,\mu$m stamp images are
in the top, middle and lower lines. The stamp size is 1.5'$\times$1.5'. The first column is the original Herschel stamp images with the MIPS 24$\,\mu$m contours (Green)
over-plotted on them and the SMG position is marked with a red circle. The second column is the best fit model image. The third
column is the stamp image only for the SMG counterpart after subtracting all neighboring objects. The Fourth column shows the
background images after subtracting all objects. \label{f:stamp1}}
\end{figure}
\begin{figure}
\includegraphics[height=160mm, width=120mm,angle=-90]{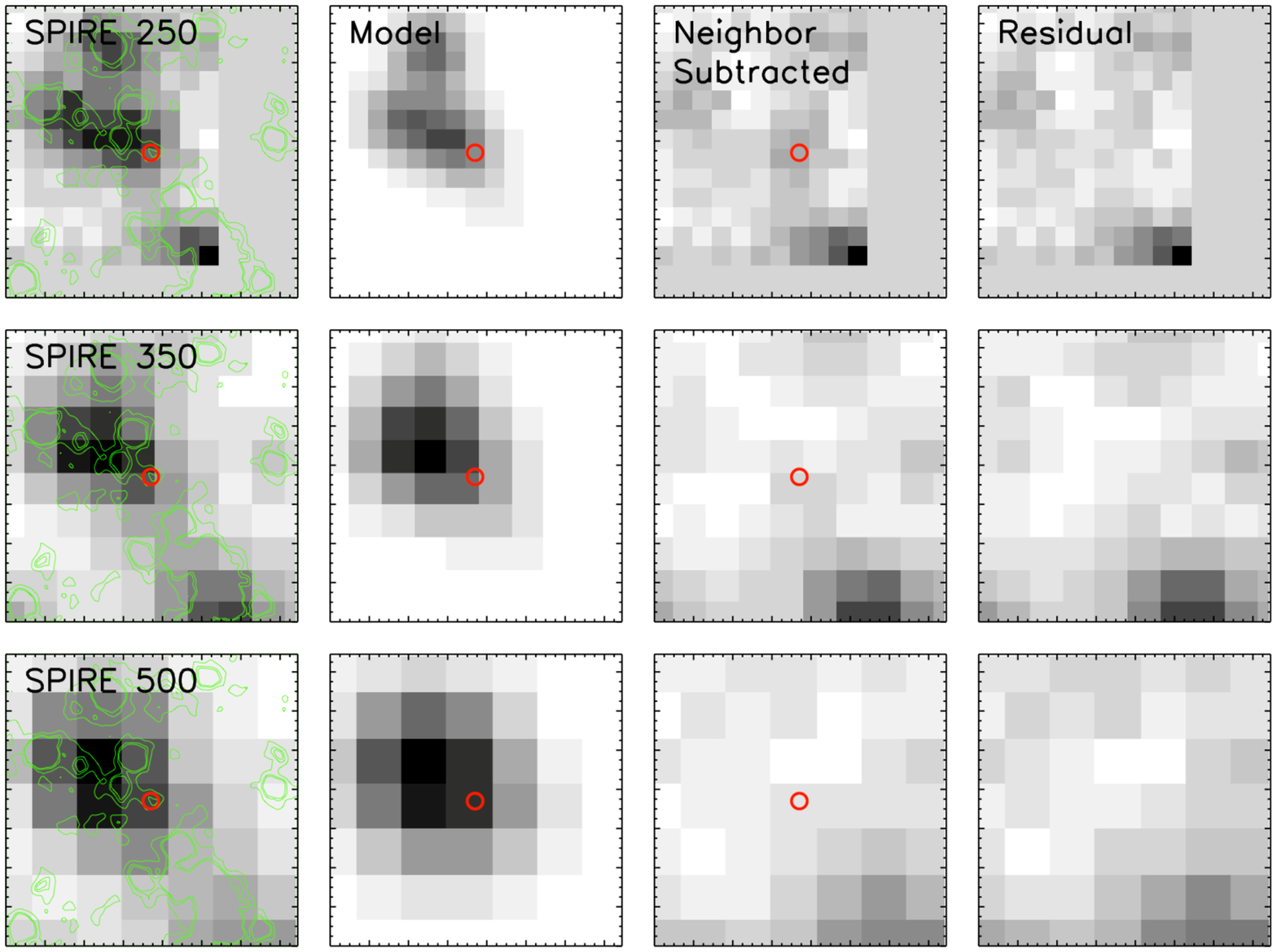}
\caption{The SPIRE stamp images of GN20.2.  \label{f:stamp2}}
\end{figure}
\begin{figure}

\includegraphics[height=160mm, width=120mm,angle=-90]{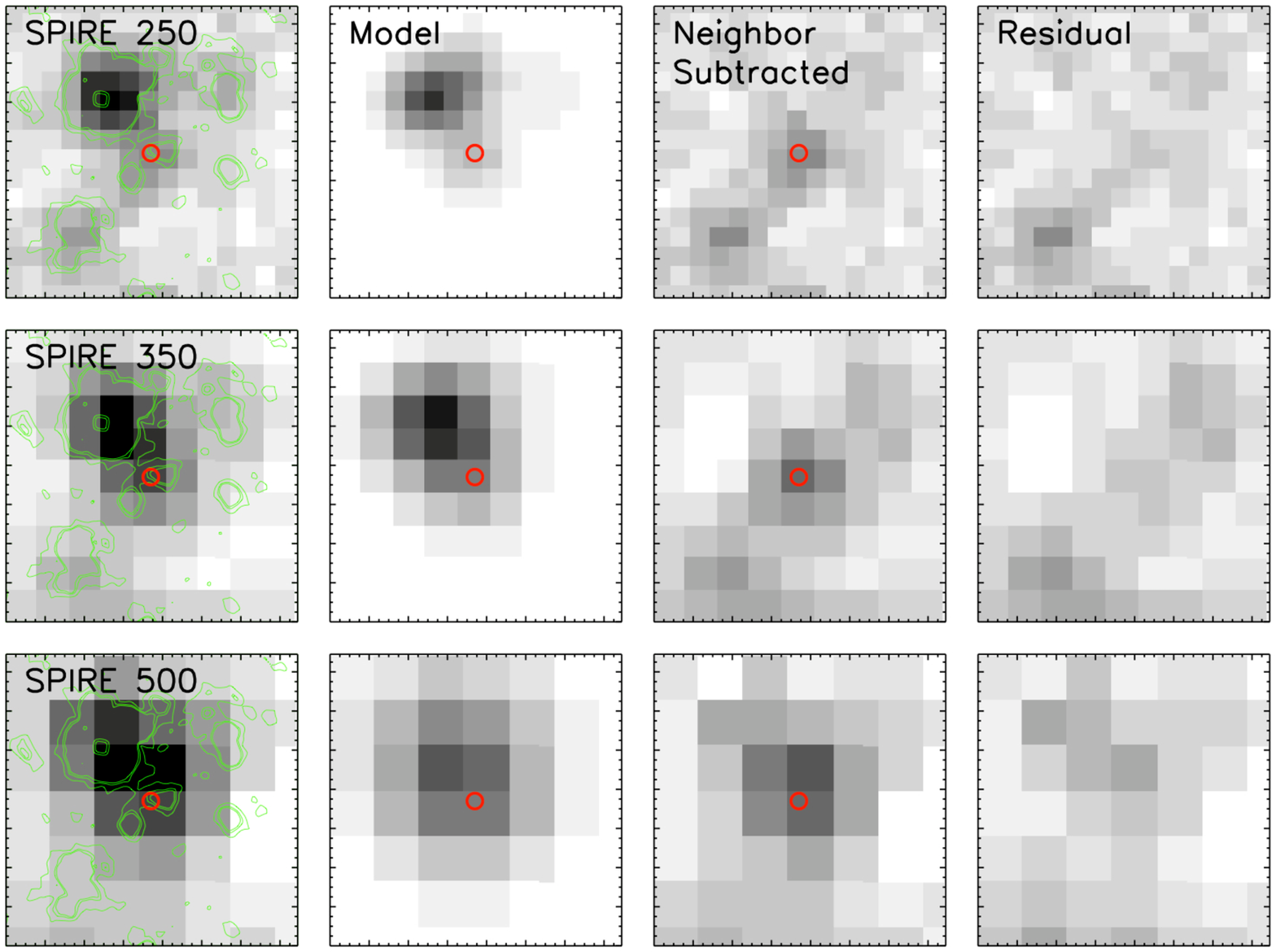}
\caption{The SPIRE stamp images of GN10.  \label{f:stamp3}}
\end{figure}
\begin{figure}

\includegraphics[height=160mm, width=120mm,angle=-90]{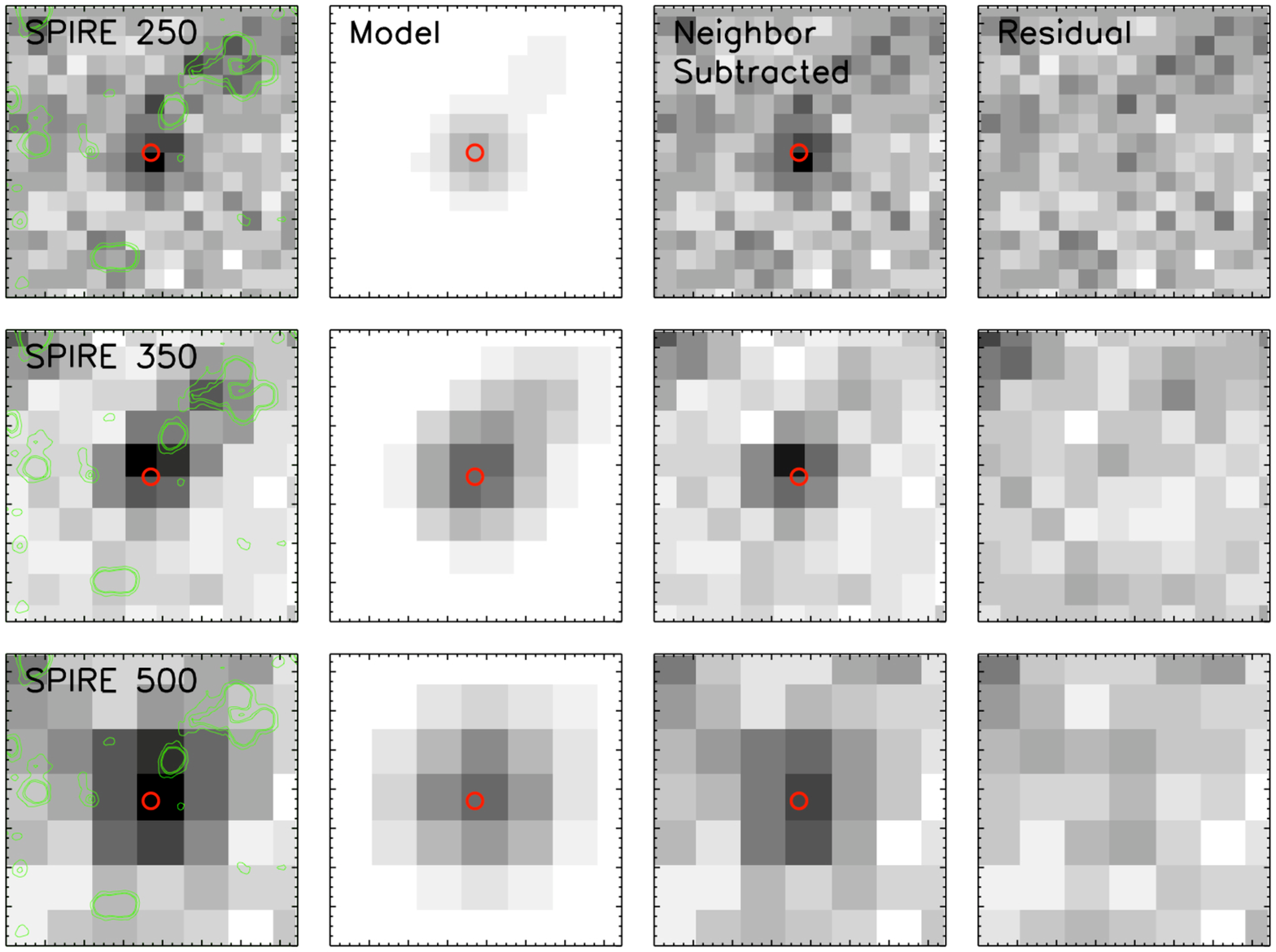}
\caption{The SPIRE stamp images of AzTEC1.  \label{f:stamp4}}
\end{figure}
\begin{figure}

\includegraphics[height=160mm, width=120mm,angle=-90]{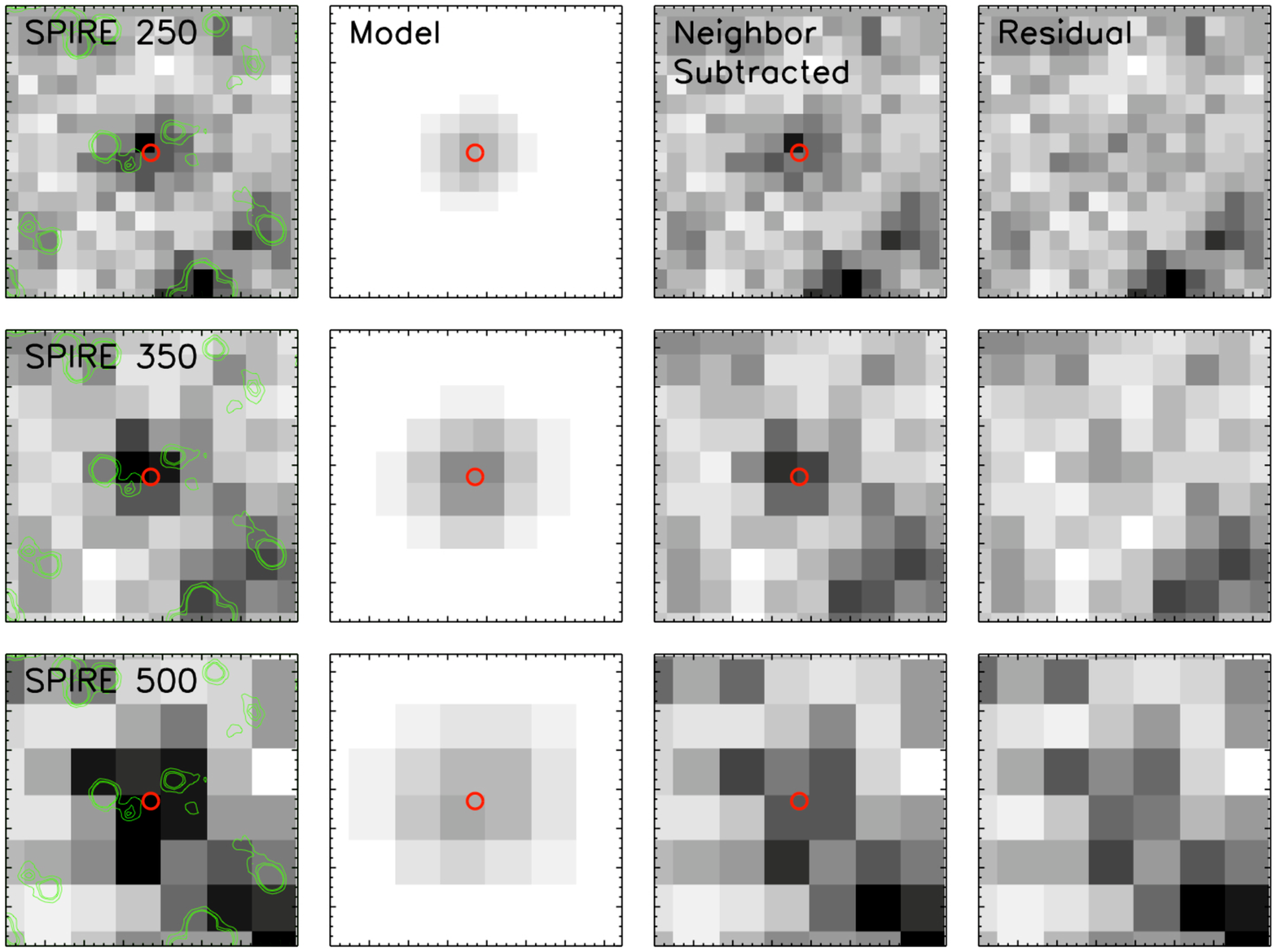}
\caption{The SPIRE stamp images of AzTEC3.  \label{f:stamp5}}
\end{figure}
\begin{figure}

\includegraphics[height=160mm, width=120mm,angle=-90]{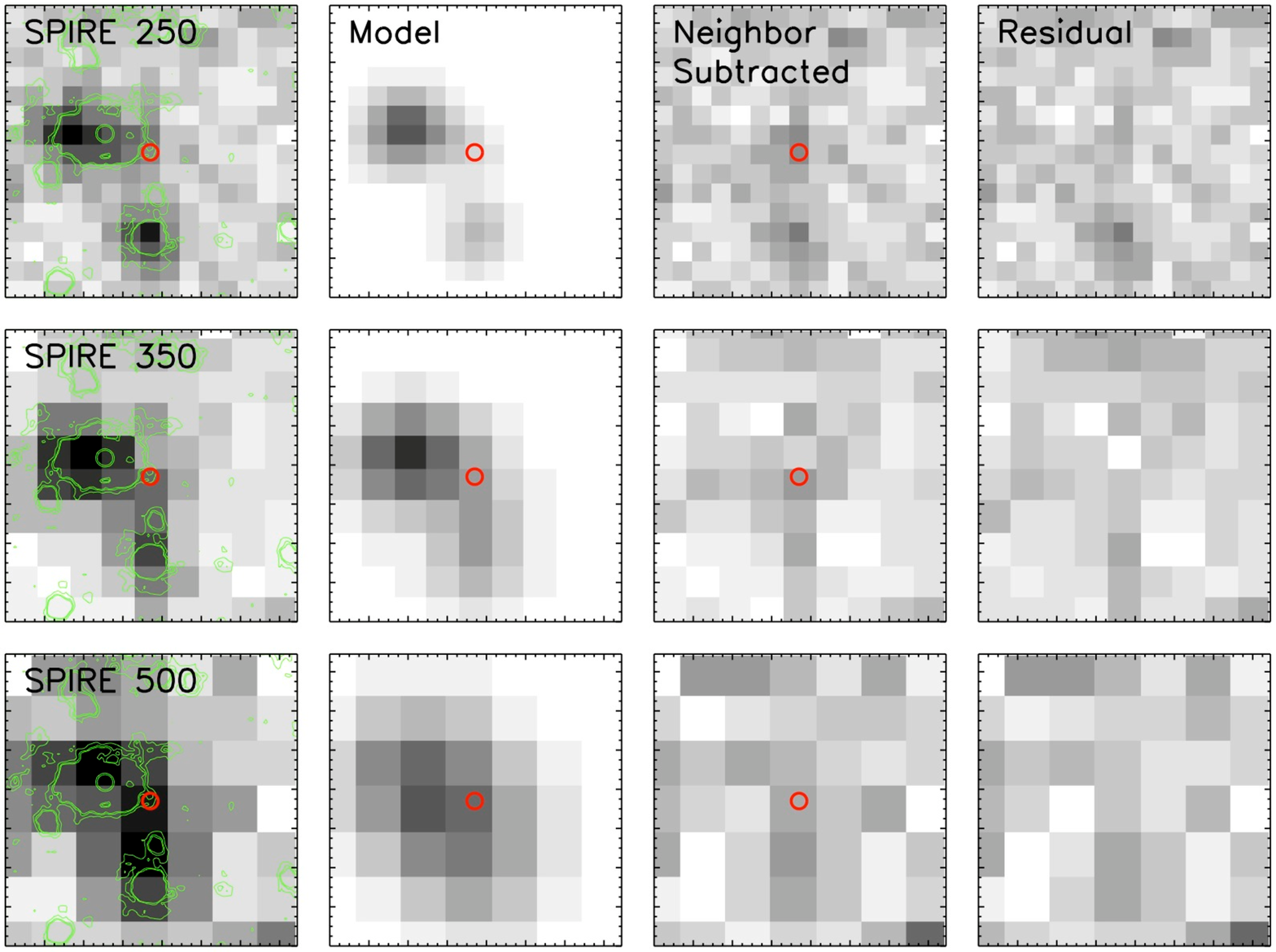}
\caption{The SPIRE stamp images of Capak4.55.  \label{f:stamp6}}
\end{figure}

The SPIRE imaging for GOODS-N, GOODS-S and COSMOS are amongst the deepest observations in HerMES. 
The 1$\sigma$ confusion level is 6mJy for all three SPIRE bands, and the
confusion limit is  $\sim$10, 12, and 20mJy at 250, 350 and 500$\,\mu$m \citep{nguyen2010}. First we performed a visual inspection of the SPIRE images. Six SMGs appear to be detected by SPIRE, while no detection is found for LESS J033229.4-275619 in the ECDFS.
The SPIRE stamp images are shown in Figures~\ref{f:stamp1}-\ref{f:stamp6}. 
Subsequent analysis to break the confusion and extract accurate flux densities for each source is thus necessary. 

Blind source extraction results in single-band catalogs \citep{smith2010} but is not sensitive to very faint sources. Prior-based catalogs \citep{roseboom2010} use MIPS 24$\,\mu$m source positions to guide the SPIRE photometry.
The HerMES team has adopted this method to produce multi-wavelength catalogs, the so-called XID catalogs \citep{roseboom2010}.
Depth and photometric accuracy for the XID catalogs critically depends on depth of the input MIPS 24$\,\mu$m catalogs \citep{roseboom2010}.
Several studies of high redshift SMGs use SPIRE photometry based on the HerMES XID catalogs
\citep{magdis2011,magdis2012,magnelli2012}. We argue that the existing HerMES XID catalogs in the COSMOS and GOODS-N fields are, however,
not optimized for our sample. First, the MIPS 24$\,\mu$m catalog used for the XID catalog in the GOODS-N field is much deeper
than that used in the COSMOS field, which is likely to cause systematic differences in the SPIRE photometry for the sources
in each fields. Second, some SMGs have either
very weak or absent MIPS 24$\,\mu$m detections, therefore they are not included in the XID catalogs. For example AzTEC1 and
AzTEC3 are not detected at 24$\,\mu$m in SCOMOS\citep{sanders2007},
however, the submillimeter/millimeter observations can be used to guide source identification in the  Herschel SPIRE bands for these sources. 
We took advantage of our own ultra-deep MIPS 24$\,\mu$m images for the sources in the COSMOS field and the FIDELS MIPS
24$\,\mu$m image in the GOODS-N field, and developed a very similar
source extraction and photometry method to the one described in \citet{roseboom2010}, but using submillimeter positions as priors for the sources of interest ($z >$ 4 SMGs) and MIPS 24$\,\mu$m positions for their neighbouring sources.
Our method proceeds as follows:
We first model the Herschel flux distribution for each SMG and its unresolved neighboring objects in each stamp image. We use the SPIRE Point-Spread-Function (PSF) 
to re-construct the flux distribution in each stamp image with the SMA or radio position for each SMG and 
MIPS 24$\,\mu$m positions for the neighboring objects.  We fit the model image to each observed stamp image with the flux density of each model object as free input 
parameter. The flux density for each object is then found from the best fits. 
 Figure~\ref{f:stamp1}-\ref{f:stamp6} show the source extraction process for each object in the present sample.  Our method yields very similar photometry for GN20  as that in \citet{magdis2011} and in \citet{roseboom2012}. Yet our 500$\,\mu$m flux density for GN20 is lower than  in \citet{magdis2011}.
\citet{roseboom2010} found that the XID catalogs may under-estimate
500$\,\mu$m photometry for high-redshift sources by assigning their flux densities to their neighbors. Our photometry 
in 500$\,\mu$m band may suffer the same problem, such a bias becomes worse when there are a number of neighboring sources in deep MIPS 24$\,\mu$m images.

\begin{figure}
\epsscale{.60}
\plotone{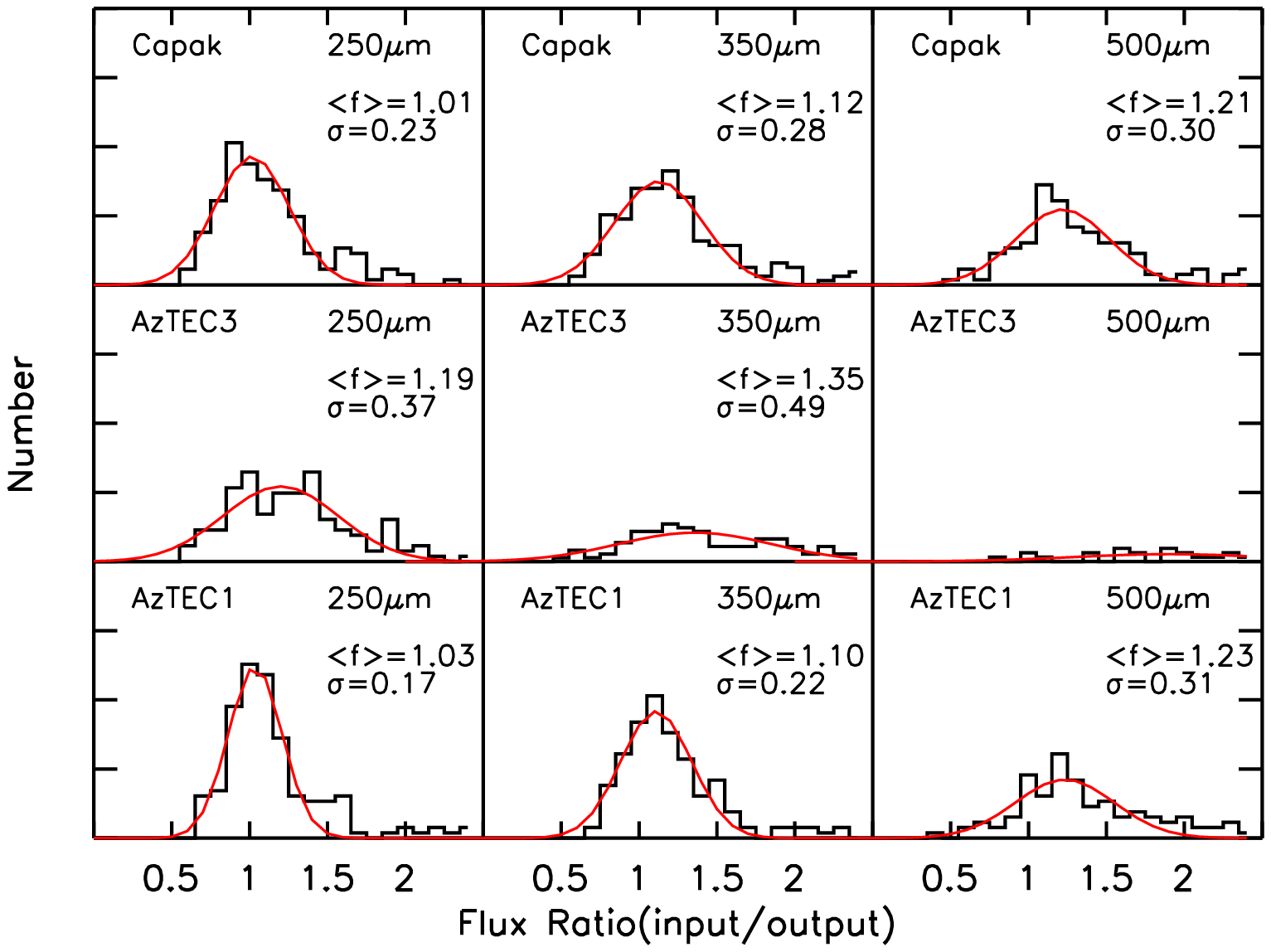}
\plotone{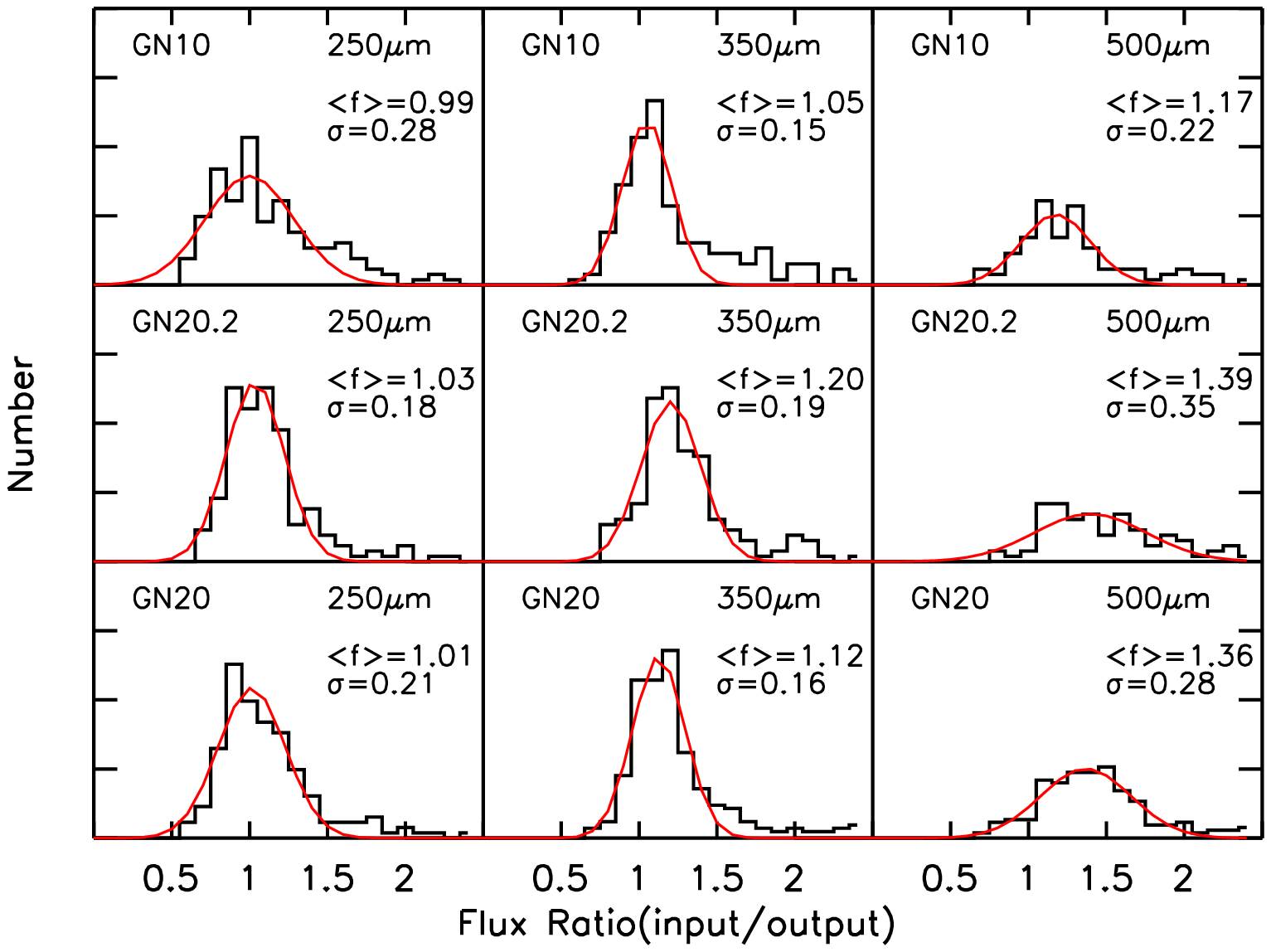}
\caption{The histogram of the recovered flux density ratio in the Monte-Carlo simulation. Both plots show how reliable our photometry
is in each SPIRE bands. The mean-flux-density-sigma ratio, $<$f$>$/$\sigma$, for each histogram measures significance of the mean recovered flux density. For example, $<$f$>$/$\sigma <$ 3 means that the mean recovered flux density is smaller than 3$\sigma$. The recovered flux density ratio histograms for all 500$\,\mu$m images show a tail at the lower end and very low $<$f$>$/$\sigma$ ratio, indicating
that their measured flux densities are very likely underestimated due to neighboring object overlapping.  Following this results, we chose not to use the 500$\,\mu$m flux densities in the following analysis.\label{f:simu}}
\end{figure}

We asses the reliability of our SPIRE photometry through Monte-Carlo simulations.
The photometry derived for each target is critically dependent on its neighbor object distribution.  Because our fitting is based on well-determined SMA or MIPS 24$\,\mu$m positions, 
source positions are fixed in the simulation.  We only randomized the flux densities 
and background counts according to their measured standard deviation.   A total of 1000 simulated stamp images were created for each target in each SPIRE band. 
We then run our photometry software on each of them and compare the measured flux densities with the input ones.  In Figure~\ref{f:simu}, we show the histogram of flux ratio between input and measured flux density in the simulation for all six objects at 250, 350 and 500$\,\mu$m.  The mean flux ratio of $f_{\rm input}$/$f_{\rm measured}$ indicates a possible bias of our actual photometry.
This simulation shows that 
the distribution of $f_{\rm input}$/$f_{\rm measured}$ for the 250$\,\mu$m photometry is gaussian centered around unity, suggesting that the 250$\,\mu$m photometry for most sources is
very robust. The 350$\,\mu$m histogram also shows a gaussian shape centering around unity, but with a few measurements much lower than input flux densities.
We argue that the 350$\,\mu$m photometry is reasonably reliable except for AzTEC3. 
We do not apply the bias measured 
in the $f_{\rm measured}$/$f_{\rm input}$ to the 250 and 350$\,\mu$m photometry for the sources in the sample, because this bias is smaller than its standard deviation for most sources.
The $f_{\rm input}$/$f_{\rm measured}$ ratios for the simulated 500$\,\mu$m photometry show a much wide
distribution. A lot of measured flux densities are much lower than the input ones in the simulation, and their
ratios are much larger and beyond the range in Figure~\ref{f:simu}. This means that our 500$\,\mu$m photometry
may be well underestimated. Thus  the measured 500$\,\mu$m flux densities are not reliable and will not be used in the following analysis.

\section{SEDs and FIR Properties for the SMGs at $z >$ 4}

\subsection{Mid-IR SEDs}

The rest-frame near-Infrared (near-IR) SEDs can be used to infer the energy source that powers infrared galaxies.
In ULIRGs and HyperLIRGs both stellar light and AGN-heated dust light can contribute to the rest-frame NIR band. 
 Several groups \citep{houck2005, yan2007,dey2008,huang2009} used IRAC photometry to classify SEDs for ULIRGs and HyperLIRGs at $z \sim$ 2. 
There are two types of
SEDs for these IR galaxies: those dominated by  stellar emission with the 1.6$\,\mu$m bump shifting into the IRAC bands, known 
as ``bumpers'' \citep{farrah2008,huang2009}; and those having strong power-law dust emission in near-IR/mid-IR and subject to 
severe dust extinction in the optical band, known as Dust Obscured Galaxies \citep[DOGS]{houck2005,yan2007,dey2008}. Spitzer IRS spectroscopic studies show different spectra in the rest 6$\,\mu$m$ < \lambda <$ 20$\,\mu$m for bumpers and DOGs: bumpers have strong PAH emissions at 6.2, 7.7,
8.6 and 11.3$\,\mu$m; DOGs have power-law continuum with no PAH emission features but strong silicate absorption at 9.6$\,\mu$m.
The mid-IR spectroscopy of SMGs at $z \sim$ 2 \citep{lutz2005, valiante2007, pope2008, mend2009} also shows strong PAH emission features
in their mid-IR spectra. 

\begin{figure}
\epsscale{.80}
\plotone{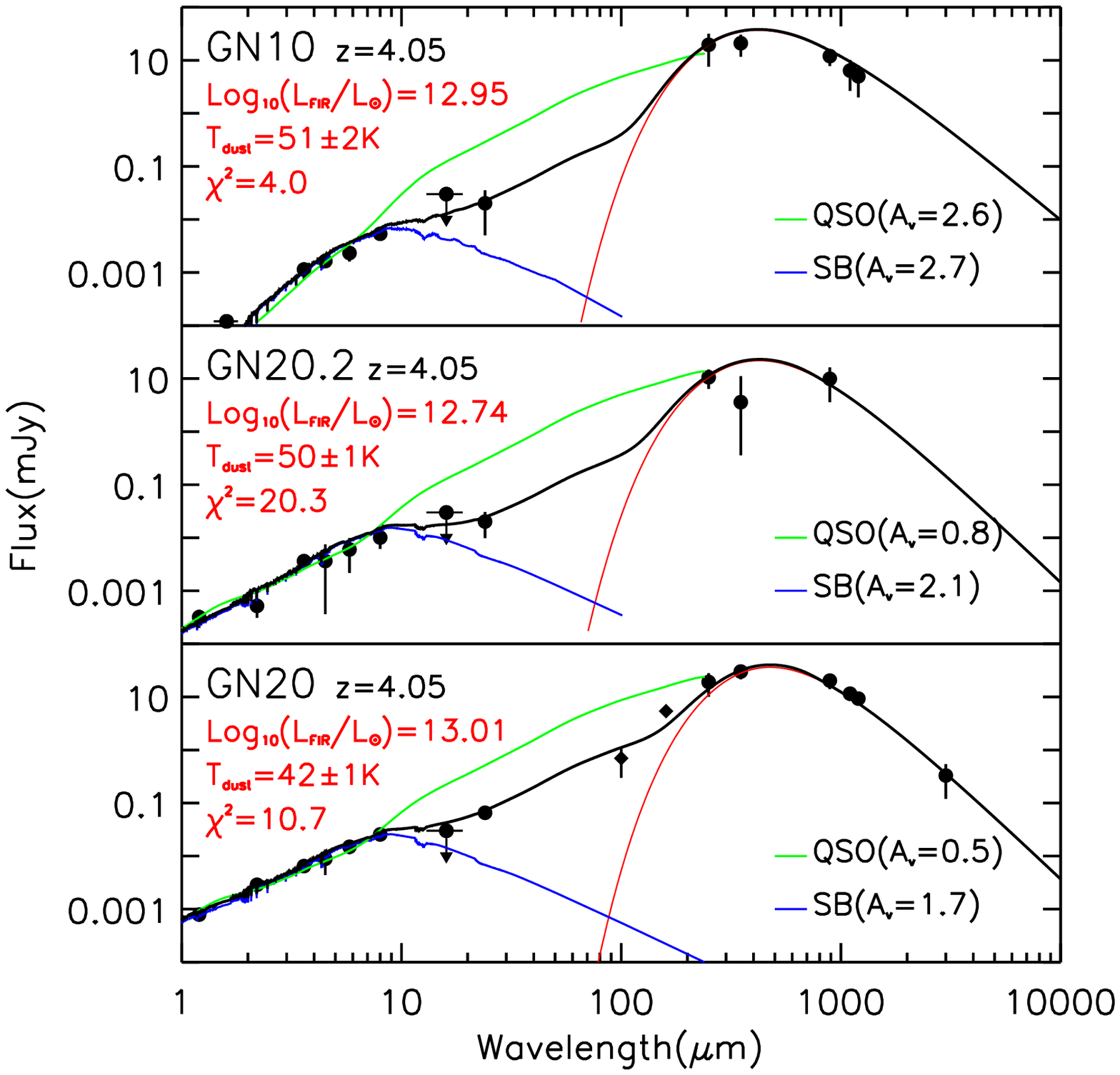}
\caption{Observed-frame SED plots for 3 SMGs at $z>$4 in the GOODS-North field. The SEDs are plotted in the wavelength range from J-band to
mm band. The green line is the dusty QSO SED template of \citet{elvis1994}. The blue line
is the dusty starburst template with a 25Myr stellar population \citep{bc03}. A$_v$ values for both templates are marked in each object panel. 
The red line is the best fitting modified Planck  function. The redshift, dust temperature and FIR luminosity for
each object are listed in each SED panel. The black line is the three-component composite SED model fitting to the full-wavelength SED of each object.\label{f:sed1}}
\end{figure}

\begin{figure}
\epsscale{.80}
\plotone{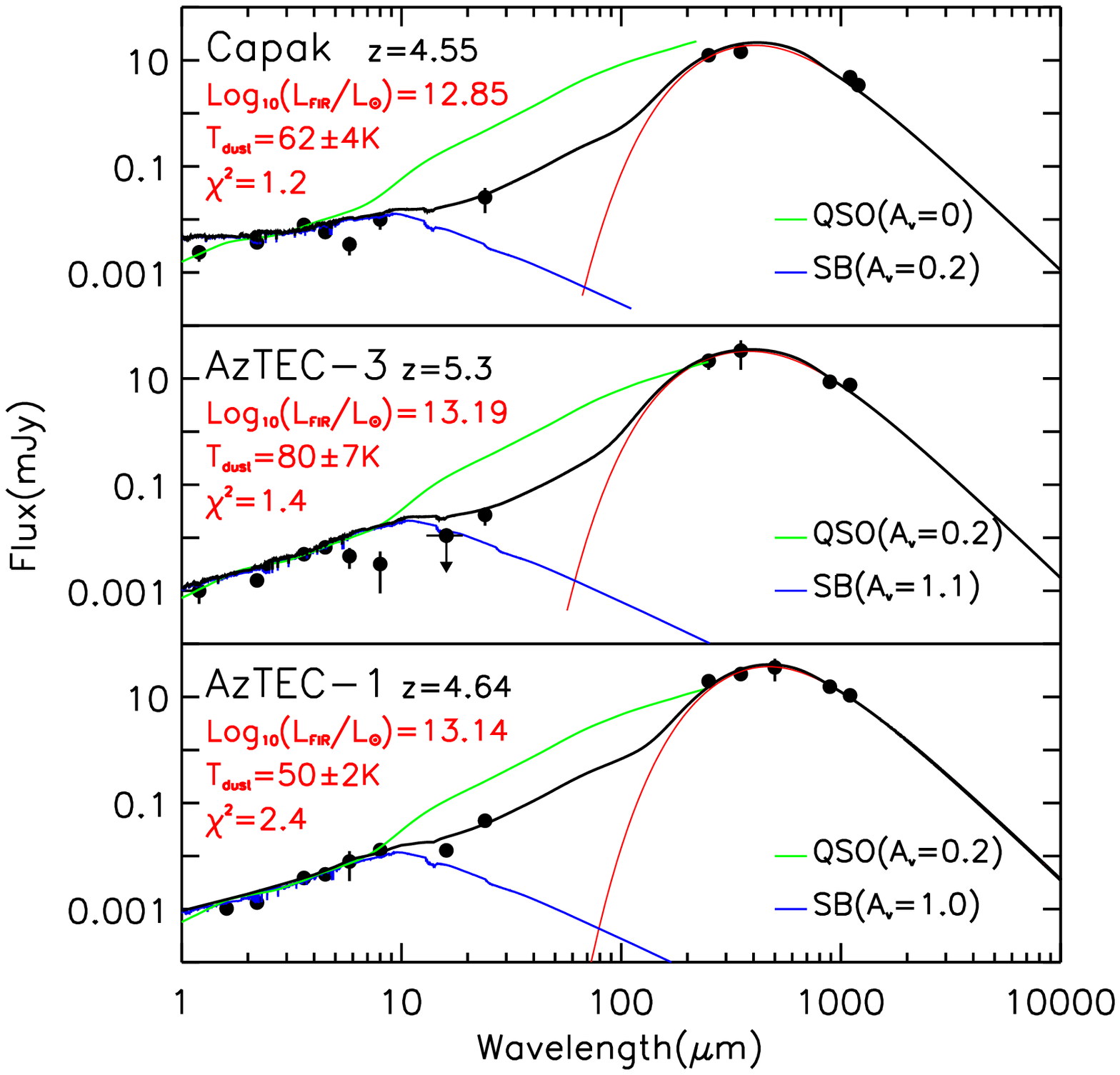}
\caption{Observed-frame SED plots for 3 SMGs at $z>$4 in the COSMOS field. The SEDs are plotted in the wavelength range from J-band to
mm band. The green line is the dusty QSO SED template of \citet{elvis1994}. The blue line
is the dusty starburst template with a 25Myr stellar population \citep{bc03}. A$_v$ values for both templates are marked in each object panel. 
The red line is the best fitting modified Planck  function. The redshift, dust temperature and FIR luminosity for
each object are listed in each panel. The black line is the three-component composite SED model fitting to the full-wavelength SED of each object.
\label{f:sed2}}
\end{figure}

At $z >$ 4 the strong PAH features at 7.7, 8.6 and 11.3$\,\mu$m shift out of the IRS effective wavelength coverage, while 
the PAH features at 3.3 and 6.2$\,\mu$m are too faint to be detected (see however Riechers et al. 2013 for a detection of weak PAH emission in GN20). We have to use the shape of the rest-frame near-IR SEDs to investigate the energy sources for SMGs at $z >$ 4.
SEDs of SMGs in the present sample show a power-law shape in the IRAC bands with [3.6] $-$ [8.0] $\sim$ 1.2 (Figure~\ref{f:sed1} and \ref{f:sed2}). Yet this does not means that they are pow-lawer objects, since the 1.6$\,\mu$m feature shifts beyond the observed 8$\,\mu$m band at z$>$4.  We need photometry at 16 and 24$\,\mu$m to classify SED of the SMGs in this sample.
``Bumpers'' at z$>$4 should have $f_{\rm 8}$/$f_{\rm 16} >$ 1. On the other hand, galaxies with power-law SEDs always have $f_{\rm 8}$/$f_{\rm 16} <$ 1. All sources in our sample are detected at 8$\,\mu$m, yet only one object, AzTEC1, was detected at 16$\,\mu$m. 
The 16$\,\mu$m limiting flux densities are used to constrain SEDs for the remaining SMGs.

The two types of templates, dusty QSO and dusty Star-Burst(SB), can fit the rest-frame optical-near-IR SEDs of the sample 
equally well (Figure~\ref{f:sed1}-\ref{f:sed2}).  
The dusty starburst template with a 25Myr old young stellar population in the BC03 model \citep{bc03} and dust extinction in range of 0.2 $<$ A$_{V} <$ 2.0 can fit all objects except GN10. 
GN10 is extremely red and not detected in the deep HST H-band imaging with H $-$ [3.6] $>$ 4.5 \citep{wang2009,huang2011}, corresponding to A$_v >$ 8 if we use the same 25Myr old young stellar population. 
But this extreme dusty model also predicts much redder [3.6] $-$ [8.0] color than what is actually observed. \citet{huang2011} found 4 additional objects with  H $-$ [3.6] $>$ 4.5 in the GOODS-S field and suggested that both dust extinction and 
presence of an old stellar population component could be responsible for such an extreme color. We fit the SED of GN10 with an 1Gyr constant star formation template and A$_V$ = 2.7. 
The star-forming template fits these SEDs in the 1$\,\mu$m $\leq \lambda \leq$ 16$\,\mu$m reasonably well (Figure~\ref{f:sed1}-\ref{f:sed2}) and their predicted 16$\,\mu$m flux densities
are consistent with the  upper limits for this sample.
The QSO template \citep{elvis1994} with a small amount of dust extinction can also fit the 1$\,\mu$m $< \lambda < $8$\,\mu$m  SEDs for AzTEC1, GN20 and GN20.2, but predicts slightly higher flux densities at 8$\,\mu$m for AzTEC3, Capak4.55 and GN10.   

Neither the dusty QSO or the SB templates can explain the MIPS 24$\,\mu$m emission 
from these objects. All objects in the sample are detected at MIPS 24$\,\mu$m with flux densities much higher than predictions based on the star-forming
template,  thus indicating the presence of hot dust emission. The dusty QSO template, however, predicts a much higher 24$\,\mu$m flux density. On the other hand, it is very challenge to constrain the host dust emission
with only 24$\,\mu$m photometry. \citep{blain2003} suggested a composite model with a modified blackbody in FIR bands and a power-law component in MIR bands to fit SEDs of IR galaxies. Serval groups \citep{kovacs2010,casey2012, dai2011} used this model to yield a good fitting to MIR-to-FIR SEDs of SMGs and Herschel sources at high redshifts. We develop a similar model for the SMGs in our sample. Our full-wavelength-range 
SED models for this sample includes three components in the infrared SEDs of
these SMGs:  stellar emission at rest $\lambda <$ 3$\,\mu$m;  a power-law component at rest $\lambda \sim$ 5$\,\mu$m; and cold dust emission at rest $\lambda \geq$ 50$\,\mu$m. Our SED model suggests that these objects are all significantly reddened in the optical bands, with $A_V$ values 
ranging from 0.25 to 3.0. We do not include any PAH emission or silicate absorption features in our SED model, because there is no available photometry for our sample in the rest 6$\,\mu$m $< \lambda <$ 20$\,\mu$m to constrain these features. The MIPS 24$\,\mu$m is the only photometry at rest $<$ 5$\,\mu$m,
to which we simply normalize the the power-law template. We discuss modeling the FIR SEDs with this power-law model and the grey-body planck function in next section.


\subsection{Dust Temperature Estimates for the SMGs at $z >$ 4}

The FIR and Submm/mm photometry available for the present sample allow us to  measure their dust emission properties. 
We fit their FIR SEDs with a grey-body Planck function $B_{\rm \lambda}$
\begin{equation}
    B_{\lambda}=(1-e^{-\tau_{\lambda}}) B_p(T_d)
\end{equation}
 and
\begin{equation}
\tau_{\lambda}=(\frac{\lambda_0}{\lambda})^{\beta}
\end{equation}
where B$_p$ is the blackbody Planck function, $T_{\rm d}$ is dust temperature and typically $\lambda_0$ = 125$\,\mu$m. The parameter $\beta$ has a typical value in 1 $< \beta <$ 2. For most SMGs in previous studies\citep{kovacs2006, yang2007}, Eq.1 can be simplified as following
\begin{equation}
    B_{\lambda}\sim \lambda^{-\beta}B_p(T_d)
\end{equation}
if their $\lambda_{\rm obs}/(1+z)/\lambda_0 >>$ 1 in Eq.1. 

A more realistic model should include multi-dust temperature systems in these SMGs. Given only a small number of photometry points available for each object in the sample, it is a fair approximation to use of a single temperature to fit their FIR SEDs. The single temperature model yields temperatures from the coldest gas in these systems. Alternatively, \citet{blain2003} proposed a composite model to fit a possible multi-temperature system  with a single temperature
component in long wavelength bands and a power-law component in short wavelength bands. We derived the dust temperatures for our sample using the model of \citet{blain2003} with  the fixed power-law index
and unfixed index in both optical thin and thick cases(Table~3). The dust temperatures derived with
the optical thick assumption are higher than those with the optical thin assumption. The fitting with
unfixed power-law index yields a much large temperature difference between the optical thick and thin
cases. We have only one photometry at 24 micron, thus are unable to constrain the power-law SEDs at the short wavelength. In this study, we adopt the model with the fixed power-law index.

The assumption on the FIR optical depth clearly yields difference in resulting dust temperatures. Several
studies confirm this difference due to the different optical depth assumptions\citep{conley2011,magdis2011,magdis2012}. Recently \citet{conley2011} studied a lensed submm galaxies at z$\sim$3 with all MIPS(24, 70, 160$\,\mu$m), SPIRE(250, 350, 500$\,\mu$m), submm, 1-3 mm detections, permitting a better constrain on its dust temperature. They found that the optical thick model provides a better
fit to its FIR SED. GN20 in our sample have 8 photometry measurement in FIR arranging from 100$\mu$ to 3mm, providing a better constrain on the optical depth in the grey-body Planck function model.

\begin{figure}
\epsscale{.80}
\plotone{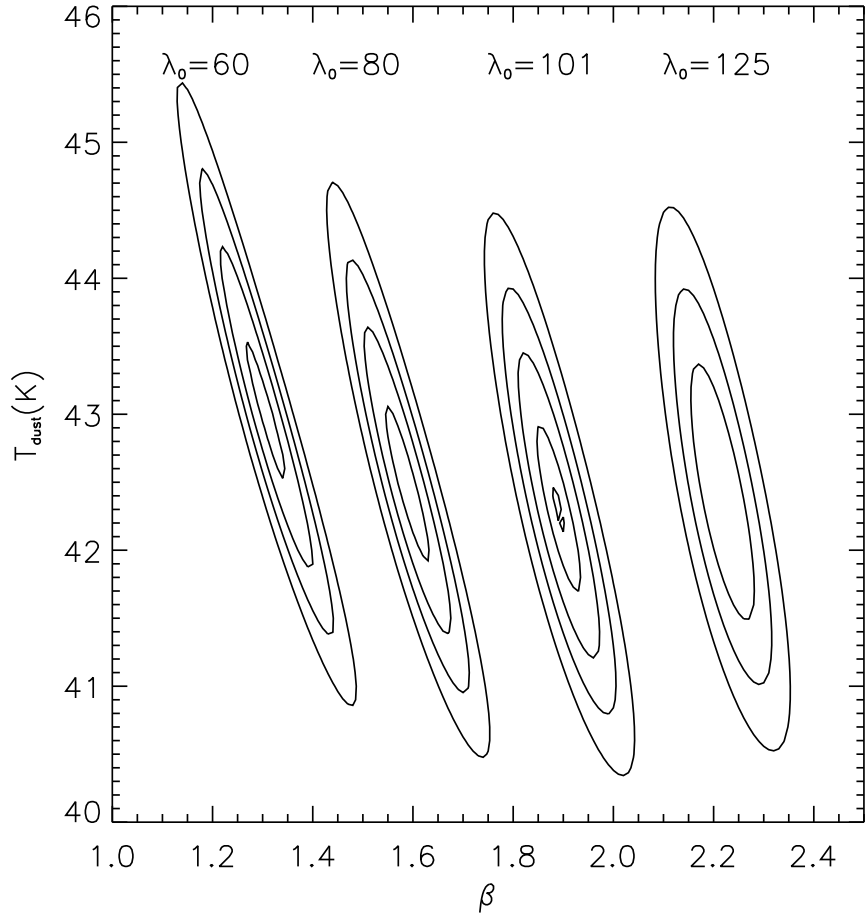}
\caption{The probability contour for the grey-body fitting to the SED of GN20, which is a three-dimension function. We over-plot 
4 projected functions in the $\beta-T_{\rm d}$ plane for $\lambda_0$ =60, 80, 101, 125$\,\mu$m. The dust temperature, however, is robust
against $\beta$ and $\lambda_0$. \label{f:con}}
\end{figure}
\begin{figure}
\epsscale{.80}
\plotone{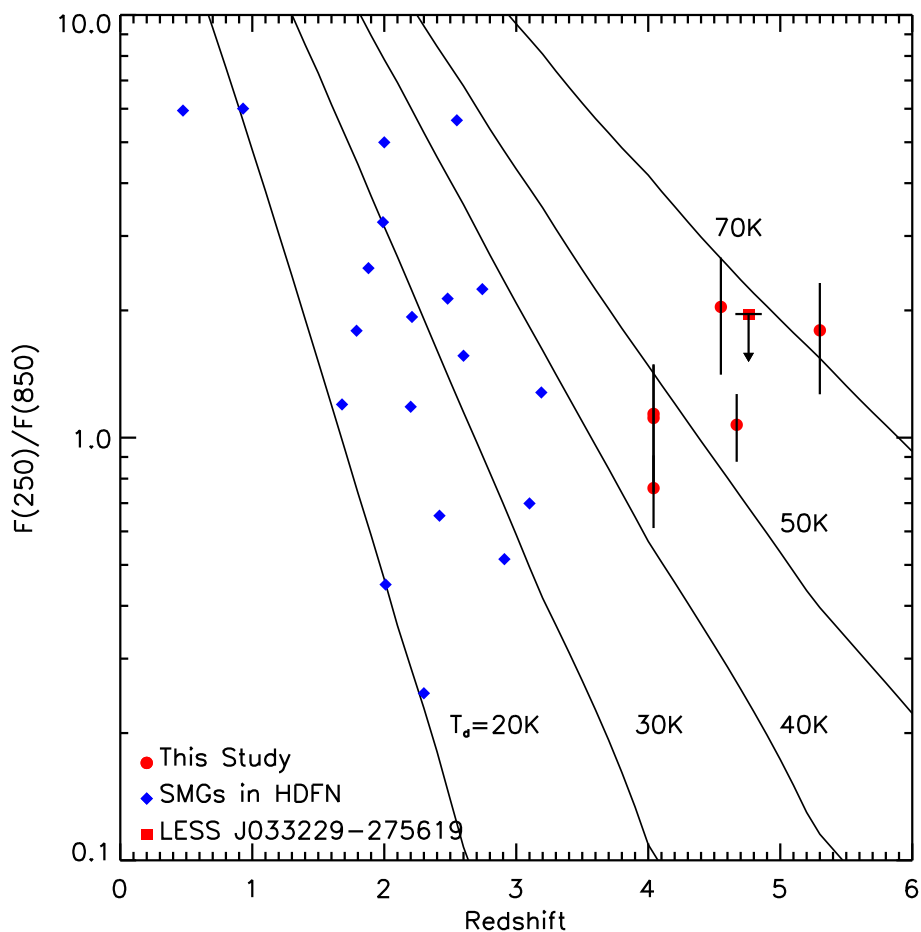}
\caption{The $F_{\rm 250}$/$F_{\rm 850}-$redshift diagram for SMGs.  The solid lines are the modified Planck  function models with $T_{\rm d}$ = 20, 30, 
40, 50 and 70 K. The submillimeter flux density at 850$\,\mu$m for Capak4.55 is extrapolated from its millimeter flux density at 1.2mm using $f_{\rm 850}$/$f_{\rm 1.2} \sim$ 2. An upper limit for LESS J033229.4-275619 is plotted in the diagram.\label{f:fr_z}}
\end{figure}

We fit the grey-body function to the SED  of GN20 with both $\beta$, $\lambda_0$ and $T_{\rm d}$ as free parameters. The best fit yields $\beta$ = 1.9, $T_{\rm d}$ = 42.5 K and  $\lambda_0$ = 101$\,\mu$m.  Figure~\ref{f:con} shows 4 slices of possibility contours as a function of $\beta$ and $T_{\rm d}$ with  $\lambda_0$ = 60, 80, 101, 125$\,\mu$m. The fit is robust in measuring $T_{\rm d}$ against both $\beta$ and $\lambda_0$, thus permitting a comparison with other objects.  \citet{magdis2011}  also fit the FIR SED of GN20 with a single temperature model, and obtain $T_{\rm d} = 32.6$ under the optically thin assumption and $T_{\rm d}$ = 46.3 under the optically thick assumption. 
Our best fit yields $\lambda_0$ = 101$\,\mu$m, thus $\tau_{100} \sim$ 1, confirming that the optically thick assumption is valid for GN20. There are only a few FIR/submm photometry data for the rest of objects which cannot constrain all parameters in the grey-body model. We argue that objects in our sample have similar FIR properties based on their similar f$_{250}$/f$_{850}$ ratio(Figure~\ref{f:fr_z}), thus adopt the single dust temperature model with $\beta$ = 2 and $\lambda_0$ = 100$\,\mu$m for all objects in the sample. The adopted dust temperatures and FIR luminosities with the optical thick assumption are reported in Table~4. 

\begin{figure}
\epsscale{.80}
\plotone{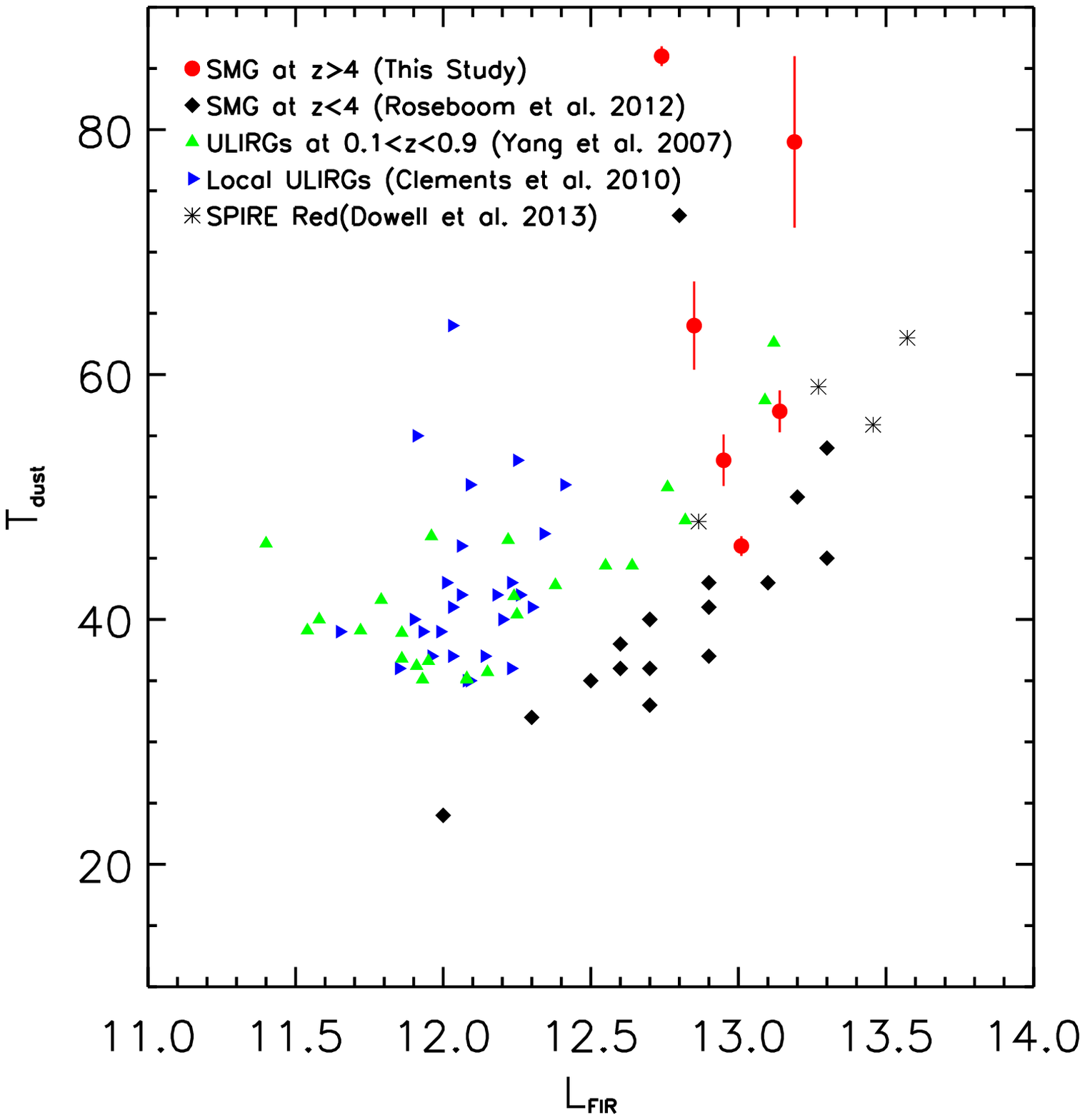}
\caption{The $L_{\rm FIR}-T_{\rm d}$ diagram for both local ULIRGs and SMGs. Dust temperatures for SMGs at $z \sim$ 2 are from \citet{roseboom2012} derived with the same method as ours. Local ULIRGs
including those at 0.1 $< z <$ 0.9 are from \citet{clements2010} and \citet{yang2007}. Objects in our sample except LESS J033229.4-275619 are plotted against local objects in this diagram.  The Herschel SPIRE red sources in the similar redshift range of 3.8$< z <$6.34 in \citet{dowell2013} are also plotted in 
this diagram for comparison, which have similar dust temperatures as objects in this sample. There is no dust temperature measurement for LESS J033229.4-275619.\label{f:fir_td}}
\end{figure}

We compare our sample with local ULIRGs and SMGs at lower redshifts in Figure~\ref{f:fir_td}. 
The three populations clearly occupy different regions in the $T_{\rm d}-L_{\rm FIR}$ diagram.
SMGs at $z \sim$ 2 generally have lower dust temperature compared to those of local ULIRGs and SMGs at $z >$ 4. \citet{kovacs2006} proposed that there exists a $L_{\rm FIR}-T_{\rm d}$ relation for SMGs. The SMGs at $z >$ 4 are at the high luminosity-high temperature 
end of the SMG distribution in Figure~\ref{f:fir_td}. Lower dust temperature for SMGs at $z \sim$ 2 is likely due to a selection
effect.  At $z \sim$ 2, the 850$\,\mu$m band samples the rest-frame $\sim$ 280$\,\mu$m, and thus, the sensitive to cold dust emission. 
In contrast, local ULIRGs selected based on IRAS 60$\,\mu$m are biased towards systems with hotter dust temperatures. On the other hand, \citet{magdis2010} measured dust temperatures for
a  MIPS 24$\,\mu$m selected ULIRG sample at $z \sim$ 1.9 \citep{huang2009}, independent of any FIR selection. 
The dust temperatures for this 24$\,\mu$m-selected sample showed no $T_{\rm d}$-bias, but cover a full range of 20 K $< T_{\rm d} <$ 60 K.   
At $z > 4$, the 850$\,\mu$m band samples the rest-frame wavelength shorter than $\sim$ 170$\,\mu$m and, preferentially  picks up SMGs with higher dust temperature than 
those at lower redshifts.  The fact that 6 out of 7 SMGs in our sample are detected at 250$\,\mu$m confirms that SMGs at z$>$4 have higher dust temperatures due to the 850$\,\mu$m selection at high redshifts.
It may also be true that those SMGs at higher redshifts may have intrinsic
higher dust temperatures. Recently \citet{Riechers2013b} and \citet{dowell2013} found that Herschel SPIRE red sources in 3.8$<$z$<$6.34 also show
to have high dust temperatures, consistent with our sample.
  
In Figure~\ref{f:fr_z}, we plot the 250-to-850$\,\mu$m flux density ratio as a function
of dust temperature and redshift based on the simple modified blackbody model described earlier. 
A typical SMG at $z >$ 4 with a single $T_{\rm d} \sim$ 30 K  and $f_{\rm 850} \sim$ 8\,mJy are too faint at 250$\,\mu$m to be detected in HerMES. Only one SMG at $z = 4.76$, LESS J033229, is not detected at 250$\,\mu$m. Its 850$\,\mu$m flux density is only 5mJy, probably too low to have a strong constraint on the dust temperature with its $f_{\rm 250}$/$f_{\rm 850}$.   


\subsection{AGN Activity in the Sample} 

  Local HyperLIRGs at $z \leq$ 1 usually harbor an AGN. The SMGs in our sample have $L_{\rm FIR} \sim$ 10$^{13}$ L$_{\odot}$. Here we performed a multi-wavelength search for signs of AGN activity in these
SMGs employing X-ray, mid-IR and radio observations. 

X-ray observations offer a direct way of identifying an AGN. The X-ray imaging in the COSMOS field is too
shallow to detect any AGN at high redshifts. X-ray imaging in GOODS-N is very deep with an exposure time of 2Ms, 
but only GN10 is detected \citep{laird2010} with
$L_{\rm x}$ = 10$^{42.93}$ erg s$^{-1}$, which corresponds to a star formation rate of 
1700 M$_{\odot}$ yr$^{-1}$ or $L_{\rm FIR} \sim$ 10$^{12.8}$ L$_{\odot}$. \citet{laird2010} argued that
intensive star formation in GN10 accounts for both the X-rays and FIR luminosities. 
On the other hand, GN20 and GN20.2 with a similar $L_{\rm FIR}$ as GN10, are not detected in the Chandra 2Ms imaging, and thus they have 
$L_{\rm x} <$ 10$^{42.6}$ erg s$^{-1}$.  If the X-ray emission from GN10 were due to intensive star formation, we would expect a similar L$_{x}$ from GN20 and GN20.2.
Both \citet{wang2009} and \citet{huang2011} suggested that there is a significant number of old stars in GN10, indicating a late stage of merging. Thus, it is
very likely that its central black accretion is switching-on in GN10, but not yet in GN20 and GN20.2. 
The X-ray luminosity for GN10 is much lower
than 10$^{44}$ erg s$^{-1}$, suggesting that it has not yet entered in the QSO phase.

\begin{figure}
\epsscale{.80}
\plotone{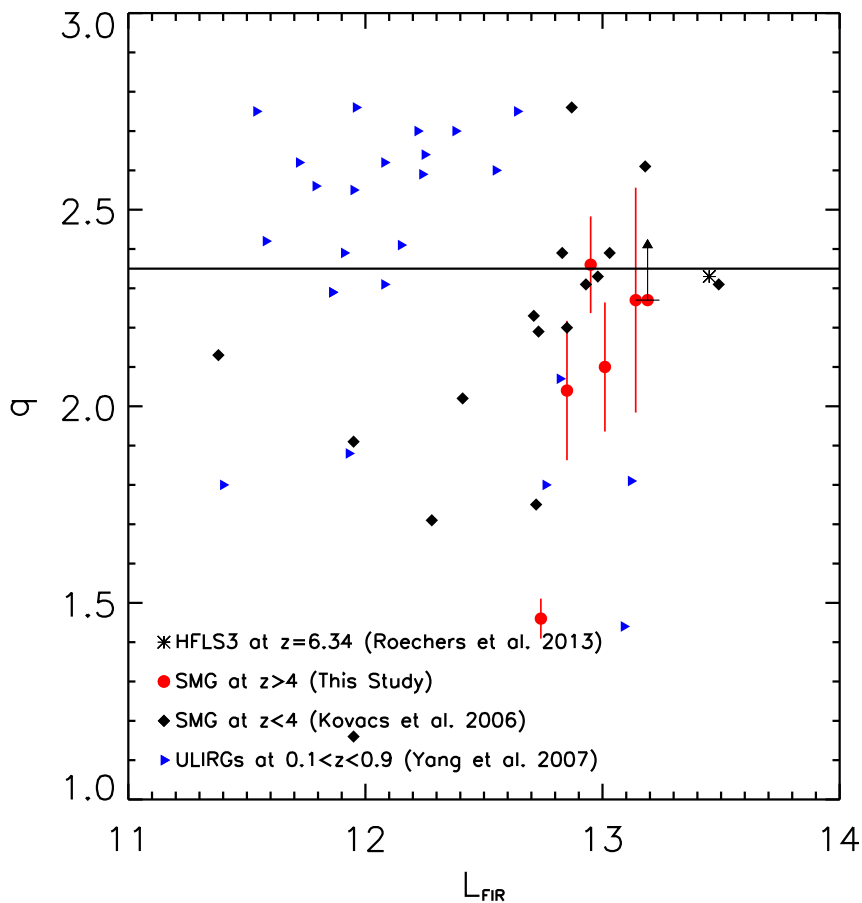}
\caption{The FIR-radio ratio q values are plotted against $L_{\rm FIR}$ for SMGs. AzTEC3 is not detected at 1.4GHz, a low-limit of q for this object is plotted. Most SMGs in this study have $q \sim$ 2.2, while GN20.2 has a much lower $q$ as $q = 1.56$, suggesting a strong AGN component. Both $L_{\rm FIR}$ and $f_{\rm 1.4GHz}$ for LESS J033229.4-275619 are from \citet{coppin2009,coppin2010}. An Extreme starburst galaxy at z=6.34 \citep{Riechers2013b} also has q=2.33.\label{f:fir_q}}
\end{figure}

The radio emission can be also used to identify AGN in FIR selected galaxies. The radio emission powered
by star formation is strongly correlated with FIR emission with radio-to-FIR flux density ratio around $q = 2.35$ \footnote{$q = Log(\frac{F_{\rm FIR}}{3.79 \times 10^{12} W m^{-2}})-Log(\frac{F_{\rm 1.4GHz}}{W m^{-2} Hz^{-1}})$, \citet{condon1982}}.  For radio loud AGN, $q$ decreases significantly due to additional non-thermal radio emission contribution from their central
black hole. Based on 350$\,\mu$m photometry \citet{kovacs2006} derived FIR luminosities for their SMG sample at $z \sim$ 2
and a radio-to-FIR flux density  ratio of $< q >$ = 2.14 $\pm$ 0.07. 
We measured a $q$ value of $\sim$2.2 for all but GN20.2 SMGs in our sample. This value of $q$ is similar to that of 
SMGs at $z \sim$ 2 (Figure~\ref{f:fir_q}). Recently \citet{Riechers2013b} found a HyperLIRG at z=6.34 with q=2.33, implying such a extreme star-burst system existing at very high redshifts without AGN signature. GN20.2 has a much lower $q$ value indicating a strong radio excess, in agreement with Daddi et al. (2009). 
All HyperLIRGs at $z <$ 1 in \citet{yang2007}'s sample have much lower $q$.  We conclude that the majority of the SMGs at $z >$ 4 (except GN20.2),  do not have significant non-thermal contribution in their radio emission, thus there is no evidence for the presence of an AGN.

All SMGs in our sample have an excess in addition to stellar photosphere emission in the rest-frame mid-IR bands (4$\,\mu$m$\ < \lambda <$ 5$\,\mu$m). It is generally thought that the mid-IR excess emission originates from hot dust heated by an AGN. However, Spitzer IRAC imaging of local star forming regions in our galaxy \citep{allen2004} and local star-burst galaxies \citep{foster2001,engelbracht2006}, also shows an excess in the continuum emission at 3.6, 4.5 and 5.8$\,\mu$m. \citet{huang2007} found
that star forming galaxies with high 8$\,\mu$m luminosities have ([3.6] $-$ [4.5])$_{\rm vega} >$ 0 and the ([3.6] $-$ [4.5])$_{\rm vega}$ colors are correlated with their 8$\,\mu$m luminosities. This indicates that galaxies with high star formation rates may display strong continuum dust emission in the mid-IR bands. The objects in our sample have extremely high star formation rates, thus we need to quantify 
their $L_{\rm 4.5}$ and compare it with other types of objects to understand its origin. 

\begin{figure}
\epsscale{.80}
\plotone{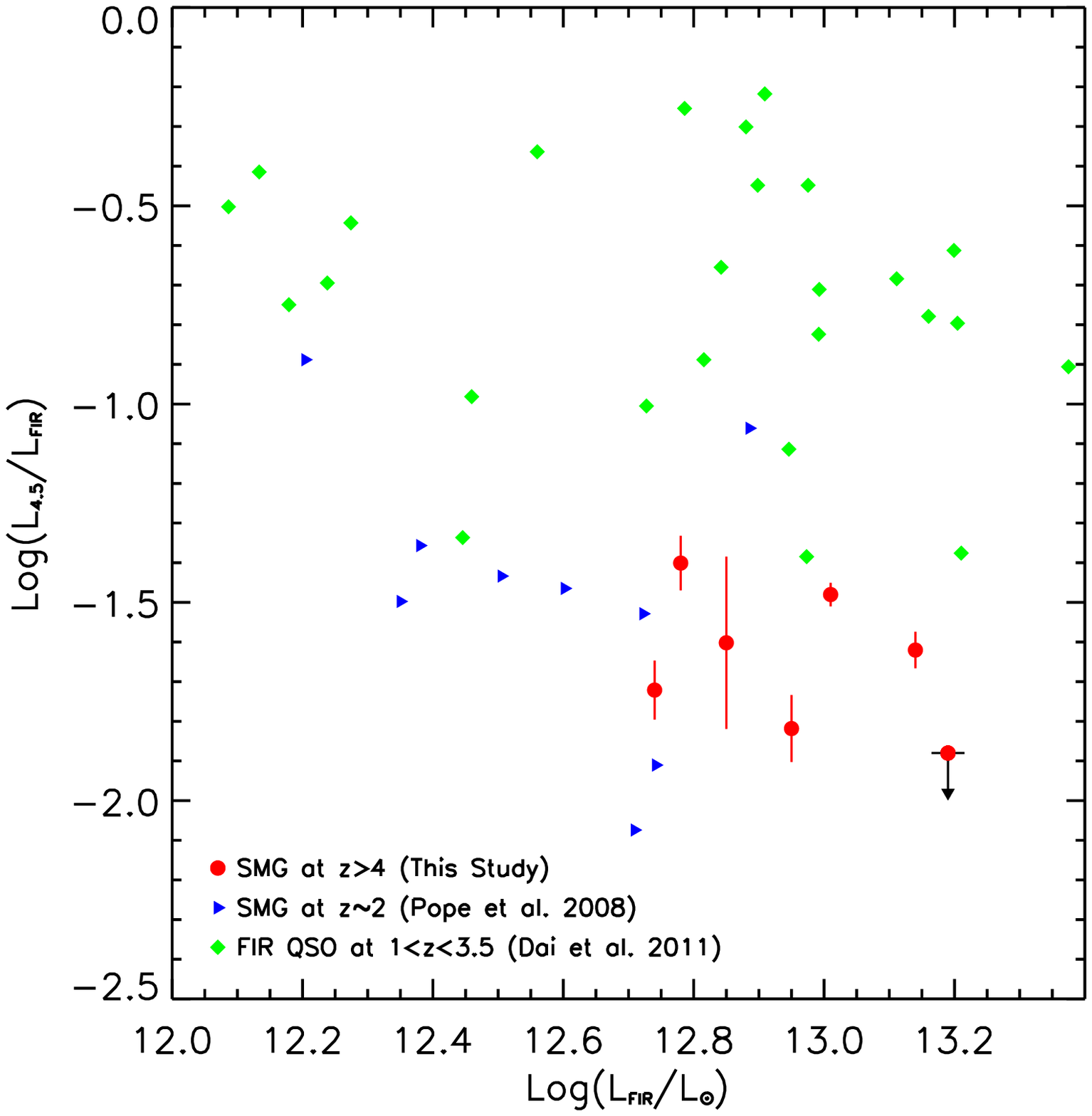}
\caption{The L$_{\rm 4.5}$/$L_{\rm FIR}$ ratio for SMGs in the GOODS-North region. SMGs at $z \sim$ 2 are from \citet{pope2008} and their 4.5$\,\mu$m luminosities, $L_{\rm 4.5}$, are calculated with their 16$\,\mu$m flux densities from the IRS peakup imaging. The 4.5$\,\mu$m luminosities for the SMGs at $z >$ 4 are calculated with their MIPS 24$\,\mu$m flux densities. Both 16$\,\mu$m band for SMGs
at $z \sim$ 2 and 24$\,\mu$m for SMGs at $z >$ 4 probe rest-frame 4.5$\,\mu$m, minimizing uncertainties caused by the K-correction variation. 
LESS J033229.4-275619 is not plotted in this diagram.\label{f:l45}}
\end{figure}

The 4.5$\,\mu$m luminosity, $L_{\rm 4.5}$, for the SMGs in this sample was calculated based on
their observed MIPS 24$\,\mu$m flux densities. The MIPS 24$\,\mu$m roughly probes the rest 4.5$\,\mu$m for the SMGs at $z >$ 4, thus it is robust in calculating 
$L_{\rm 4.5}$ for this sample with their MIPS 24$\,\mu$m flux densities. 
We also calculated L$_{4.5}$ for QSOs and SMGs at $z\sim$ 2 for comparison.
Using MIPS 24$\,\mu$m flux densities to calculate $L_{\rm 4.5}$ 
for SMGs at $z \sim$ 2, which probes the rest-frame 8$\,\mu$m emission, may introduce a large uncertainty.
The IRS spectra of SMGs at $z \sim$ 2 show many predominant features such as PAHs emission features at 6.2, 7.7 and 8.6$\,\mu$m and the silicate absorption line \citep{pope2008,mend2009}.  Variation of these features in their spectra may cause a large error for the MIPS 24$\,\mu$m K-correction in calculating 
$L_{\rm 4.5}$ for SMGs at $z \sim$ 2. Fortunately, IRS peak-up 16$\,\mu$m images are available in the GOODS-N field,  probing the rest-frame 4-5$\,\mu$m band.
SMGs at $z \sim$ 2 in this field are all detected at 16$\,\mu$m\citep{pope2008}.  Thus, their $L_{\rm 4.5}$ can be determined robustly with the measured 16$\,\mu$m flux densities \citep{pope2008, huang2009}. The selected QSOs for comparison are in 1 $< z <$ 3  and 
detected by Herschel at 250$\,\mu$m, qualifying them to be HyperLIRGs \citep{hatzi2010,dai2011}. These QSOs have much higher MIPS 24$\,\mu$m flux density and typically have
power-law SEDs in the mid-IR bands. Their $L_{\rm 4.5}$ can be easily derived using their MIPS 24$\,\mu$m flux densities\citep{dai2011}.
Figure~\ref{f:l45} shows a comparison of the $L_{\rm 4.5}$ between SMGs in this sample with the SMGs at $z \sim$ 2 and QSOs in \citet{dai2011}. The SMGs at $z \sim$ 2 show very strong PAH emission, indicating a star-burst dominated ULIRGs. 
The FIR detected QSOs have much higher $L_{\rm 4.5}$/$L_{\rm FIR}$ than SMGs at both low and
high redshifts. The SMGs at $z >$ 4 have similar $L_{\rm 4.5}$/$L_{\rm FIR}$ as SMGs at $z \sim$ 2. This comparison suggests that
the 4.5$\,\mu$m emission from SMGs at both low and high redshifts is produced by similar mechanisms, namely intense
star formation, although we cannot rule out the AGN contribution with 100\% confidence. We conclude that these SMGs with $L_{\rm FIR} \sim$ 10$^{13}$ L$_{\odot}$ are not in the QSO phase.

\subsection{Cold Gas in SMGs at $z >$ 4}
 
  SMGs experiencing intense star forming activity requires a vast cold gas reservoir. The cold gas is typically traced by molecular CO emission. All
SMGs but AzTEC-1 have CO detections, indicating that they have cold gas of 2-5 $\times$ 10$^{10}$ M$_{\odot}$ \citep{schinnerer2008,daddi2009a, daddi2009b,coppin2009,riechers2010}.
Similar amount of cold gas is found in similar types of objects: $\sim$ 3 $\times$ 10$^{10}$ M$_{\odot}$ for SMGs at $z \sim$ 2\citep{greve2005}; $\sim$ 3.4 $\times$ 10$^{10}$ M$_{\odot}$ 
for QSOs at $z \sim$ 2 \citep{coppin2008}; 1.4$-$4 $\times$ 10$^{10}$ M$_{\odot}$ for QSO at $z >$ 6 \citep{wangren2011}; and $\sim$ 1.4 $\times$ 10$^{10}$ M$_{\odot}$ for MIPS24 selected ULIRGs at $z \sim$ 2 \citep{yan2010}.  With the measurement of dust temperature, we are also able to derive the dust mass in these galaxies, their dust masses are which are 
in the range of 1 $\times$ 10$^9$ M$_{\odot} <$ $M_{\rm dust} < $6 $\times$ 10$^9$ M$_{\odot}$ (Table~3). For the derivation of the dust masses we have adopted a  dust mass absorption coefficient of $\kappa_{\rm 250} = 5.1$ cm$^2$ g$^{-1}$ (Li \& Draine 2001). 
The dust-to-gas ratio for this sample is in the range of 
$M_{\rm H_2}$/$M_{\rm dust} \sim$ 10. AzTEC-1 has a dust mass of 3.65 $\times$ 10$^9$ M$_{\odot}$, implying that it has a reservoir of $\sim$ 4 $\times$10$^{10}$ M$_{\odot}$ cold gas. 

\begin{figure}
\epsscale{.80}
\plotone{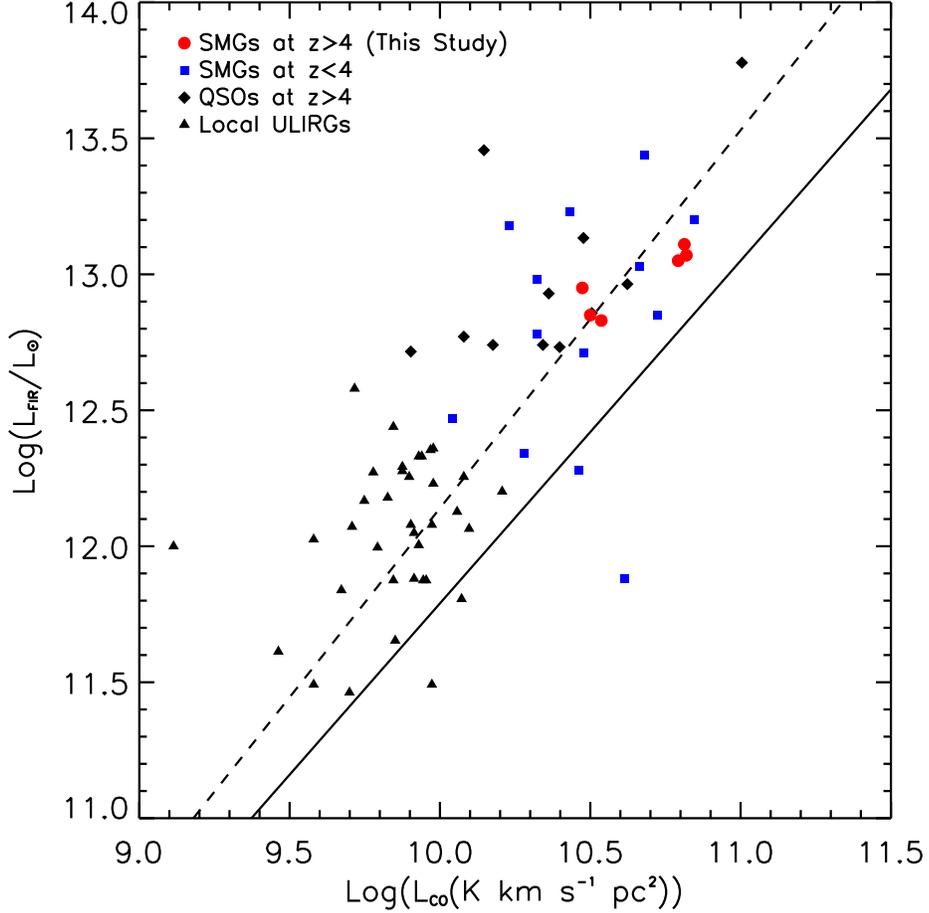}
\caption{The $L_{\rm CO}-L_{\rm FIR}$ diagram for local ULIRGs, QSOs and SMGs at various redshifts. The $L_{\rm CO}$ measurements  should be regarged as upper limits since we have assumed a line ratio of unity for sources for which only higher transition lines are available. The dashed line is the $L_{\rm CO}-L_{\rm FIR}$ relation determined with local ULIRGs and SMGs at $z \sim$ 2. The solid line is $L_{\rm CO}-L_{\rm FIR}$ relation for
local star forming galaxies. LESS J033229.4-275619 is not plotted in this diagram. \label{f:fir_co}}
\end{figure}

  We study the $L_{\rm CO}-L_{\rm FIR}$ relation for this sample with more accurate  
$L_{\rm FIR}$ and compare it with other populations. We convert CO luminosities from different J-transitions  to $L_{\rm CO[1-0]}$, assuming unity line ratio. While this assumption is not striclty correct, given the uncertainties linked with this conversion factor, it is a valid simplification in order to derive  lower limits for $L_{\rm CO[1-0]}$. We stress that this assumption  typically leads to a factor of $\sim 2-4$ underprediction of the $L_{\rm CO[1-0]}$ (e.g., Riechers et al. 2011,  Ivison et al. 2011, Carilli et al. 2010, Hodge et al. 2013). The SMGs
at $z >$ 4 are located in the same region in the $L_{\rm CO}-L_{\rm FIR}$ diagram (Figure~\ref{f:fir_co}) as SMGs at $z \sim$ 2, 
consistent with the $L_{\rm CO}-L_{\rm FIR}$ relation determined by local ULIRGs and lower 
redshift SMGs \citep{downes1998, solomon2005, daddi2010,genzel2010}. This indicates that the SMGs at $z >$ 4 have similar star formation efficiency as
those at $z \sim$ 2.

\section{Summary}

  We have performed the first multi-wavelength study of a sample of seven SMGs at $z >$ 4.0.
The mid-IR photometric data, including ultra-deep IRS peakup imaging and MIPS 24$\,\mu$m imaging for a part of this sample, were obtained 
in the  Spitzer cryogenic mission. The FIR photometry of this sample comes from the recent Herschel SPIRE survey (HerMES).
Six objects in the sample are detected in the SPIRE 250 and
350$\,\mu$m bands. We combine ground-based near-IR, Spitzer IRAC and MIPS, Herschel SPIRE and ground-based submillimeter and millimeter photometry  and obtain SEDs for this sample in the full IR wavelength range. We are able to fit the rest optical-NIR SEDs
of this sample with both dusty starburst and QSO templates. The dusty QSO templates underestimate the measured MIPS 24$\,\mu$m flux
densities for all objects in the sample. The deep 16$\,\mu$m imaging place a strong constrain on the 
origin of the optical-NIR part of the SED. We find that stellar emission with little or no contribution from hot dust can explain the SED up to the observed 16$\,\mu$m.  The dust extinction values for this sample are in the range 0.2$<$A$_v <$3.0. We suggest a three-component composite model to fit the full SED
of these objects: a stellar emission component at $\lambda_{\rm obs} <$ 16$\,\mu$m; a hot dust emission at 24$\,\mu$m; and
a cold dust emission at $\lambda_{\rm obs} >$ 250$\,\mu$m.

  At $z >$ 4, the 850$\,\mu$m band selects SMGs with higher dust temperatures and higher FIR luminosities than those at z$\sim$2. 
Our analysis shows that a typical SMG with $T_{\rm d} <$ 40 K will have too low a 250$\,\mu$m flux density to be detected 
in the HerMES survey. A high percentage of SMGs at high redshifts are
detected at 250$\,\mu$m, suggesting their high dust temperatures.
Fitting modified Planck  functions to the FIR SEDs yield dust temperature of 40K-80K and $L_{\rm FIR}$
of $\sim$ 10$^{13}$ L$_{\odot}$ for this sample. 

 We searched for the presence of AGN signatures in this sample but found  rather weak evidence. Only GN10 is detected in X-ray 
with $L_{\rm x}$ = 10$^{42.93}$ erg s$^{-1}$. GN20.2
has an excess radio emission with radio-to-FIR ratio of $q = 1.46$. All remaining objects have q $\sim$ 2.2, suggesting that
their radio emission is powered by intense star formation.  Almost all objects are detected at MIPS 24$\,\mu$m, thus have
rest-4.5$\,\mu$m luminosities of $L_{\rm 4.5} \sim$ 10$^{11.5}$L$_{\odot}$. The 4.5$\,\mu$m luminosities of these objects
are much lower than those of QSO. We suggest that these SMGs show absent or weak AGN features in X-ray, mid-IR and radio bands, but the QSO phase has not yet appeared in these SMGs. The $L_{\rm CO}-L_{\rm FIR}$ relation for this sample is also consistent with that determined for
local ULIRGs and starburst-dominated SMGs at $z \sim$ 2.  These submillimeter galaxies at $z >$ 4 are hotter and more luminous in FIR, but otherwise very similar to those at $z \sim$ 2. We conclude
that, even though these SMGs are HyperLIRGs and may harbor weak AGNs, their QSO phase has yet to be triggered.

\acknowledgments

This work is based in part on observations made with the Spitzer Space Telescope, which is operated by the Jet Propulsion Laboratory, California Institute of Technology under a contract with NASA. Support for this work was provided by NASA through an award issued by JPL/Caltech. 
Herschel is an ESA space observatory with science instrumen
ts provided by European-led Principal Investigator consortia an
d with important participation from NASA.
SPIRE has been developed by a consortium of institutes led
by Cardiff Univ. (UK) and including: Univ. Lethbridge (Canada);
NAOC (China); CEA, LAM (France); IFSI, Univ. Padua (Italy);
IAC (Spain); Stockholm Observatory (Sweden); Imperial College
London, RAL, UCL-MSSL, UKATC, Univ. Sussex (UK); and Caltech,
JPL, NHSC, Univ. Colorado (USA). This development has been
supported by national funding agencies: CSA (Canada); NAOC
(China); CEA, CNES, CNRS (France); ASI (Italy); MCINN (Spain);
SNSB (Sweden); STFC, UKSA (UK); and NASA (USA).



Facilities: \facility{Herschel(SPIRE)}, \facility{Spitzer(IRAC, MIPS)}.



\clearpage




\clearpage

\clearpage

\clearpage


\clearpage

\clearpage

\clearpage

\clearpage

\clearpage

\clearpage

\clearpage

\begin{deluxetable}{lcccc}
\tabletypesize{\scriptsize}
\tablecaption{The Submillimeter Galaxy Sample at z$>$4\label{tbl-1}}
\tablewidth{0pt}
\tablehead{
\colhead{Name} & \colhead{RA(2000)} &\colhead{DEC(2000)} &\colhead{z} &  
}
\startdata
GN10          &12:36:33.45$^{1}$&~~62:14:08.7$^{1}$& 4.05 & \citet{wang2007}\\
GN20          &12:37:11.90$^{2}$&~~62:22:12.1$^{2}$& 4.05 &\citet{daddi2009a} \\
GN20.2        &12:37:08.77$^{2}$&~~62:22:01.7$^{2}$& 4.05 &\citet{daddi2009a} \\
AzTEC1        &09:59:42.86$^{3}$& ~~02:29:38.2$^{3}$& 4.64 & \citet{younger2007}\\
AzTEC3        &10:00:20.70$^{3}$& ~~02:35:20.5$^{3}$& 5.30 &\citet{younger2007} \\
Capak4.55     &10:00:54:48$^{4}$& ~~02:34:35.9$^{4}$&4.55  &\citet{capak2008}\\
J033229  &03:32:29.30$^{5}$&$-$27:56:19.4$^{5}$& 4.76 &\citet{coppin2010} \\
\enddata
\end{deluxetable}
\clearpage


\clearpage

{\small
\begin{deluxetable}{lcccccccccc}
\tabletypesize{\scriptsize}
\tablecaption{The Multi-Wavlength Photometry for the Submillimeter Galaxy sample at $z >$ 4\label{tbl-2}}
\tablewidth{0pt}
\tablehead{
\colhead{Name} & \colhead{3.6$\,\mu$m} &
\colhead{4.5$\,\mu$m} & \colhead{5.8$\,\mu$m} & \colhead{8.0$\,\mu$m} & \colhead{24$\,\mu$m} &
\colhead{250$\,\mu$m} & \colhead{350$\,\mu$m} & \colhead{850$\,\mu$m} & \colhead{1100$\,\mu$m} & \colhead{1.4gHz}\\
\colhead{ } & \colhead{($\,\mu$Jy)} &
\colhead{($\,\mu$Jy)} & \colhead{($\,\mu$Jy)} & \colhead{($\,\mu$Jy)} & \colhead{($\,\mu$Jy)} &
\colhead{(mJy)} & \colhead{(mJy)} & \colhead{(mJy)} & \colhead{(mJy)} & \colhead{($\,\mu$Jy)}
}
\startdata
GN10           & 1.14$\pm$0.14 & 1.64$\pm$0.13 & 2.33$\pm$0.24 & 5.37$\pm$0.37 & 26.2$\pm$5.08 & 19.6$\pm$4.0 & 21.0$\pm$3.1 & 12.0$\pm$1.4 & 5.00$\pm$1.0 & 34.4$\pm$4.2\\
GN20           & 21.9$\pm$0.05 & 21.5$\pm$0.17 & 21.0$\pm$0.10 & 20.4$\pm$0.06 & 65.5$\pm$4.45 & 19.1$\pm$3.0 & 30.1$\pm$2.9 & 20.3$\pm$2.1 & 11.5$\pm$1.0 & 73.8$\pm$2.1\\
GN20.2        & 22.5$\pm$0.07 & 22.5$\pm$0.38 & 22.0$\pm$0 .23 & 21.4$\pm$0.14 & 20.2$\pm$3.45 & 10.6$\pm$1.4 & 3.6$\pm$2.5 & 9.90$\pm$2.1 & ---          & 170.$\pm$12.8\\
AzTEC1        & 3.87$\pm$0.13 & 4.53$\pm$0.23 & 7.90$\pm$4.50 & 13.0$\pm$2.88 & 46.4$\pm$4.90 & 19.8$\pm$2.9 & 27.0$\pm$4.4 & 15.6$\pm$1.1 & 10.7$\pm$1.3 & 48.0$\pm$14.0\\
AzTEC3        & 4.90$\pm$0.20 & 6.60$\pm$0.30 & 4.50$\pm$1.90 & 3.20$\pm$2.30 & 5.30$\pm$5.30 & 21.7$\pm$7.0 & 33.7$\pm$19.0 & 8.70$\pm$1.5 & 7.60$\pm$1.2 & ---\\
Capak4.55     & 7.90$\pm$0.20 & 5.80$\pm$0.40 & 3.40$\pm$1.30 & 10.0$\pm$3.60 & 26.0$\pm$13.0 & 12.5$\pm$2.0 & 14.5$\pm$3.4 & ---          & 4.80$\pm$1.5 & 45.0$\pm$9.00\\
J033229  & 2.90$\pm$0.10 & 4.00$\pm$0.10 & 6.3$\pm$0.40  & 9.20$\pm$0.40 & 32.0$\pm$5.00 & ---           & ---           & 5.00$\pm$1.4 & ---          & 24.0$\pm$6.30\\
\enddata
\end{deluxetable}
}

\clearpage

\begin{deluxetable}{lcccc}
\tabletypesize{\scriptsize}
\tablecaption{The Derived Dust Temperatures from Various Models\label{tbl-4}}
\tablewidth{0pt}
\tablehead{
\colhead{Name} & \colhead{T$_{dust}$ [K]} & \colhead{T$_{dust}$ [K]} & \colhead{T$_{dust}$ [K]} & \colhead{T$_{dust}$ [K]}\\
\colhead{~~~} & \colhead{Optical Thick/Fixed Index} & \colhead{Optical Thin/Fixed Index} & \colhead{Optical Thick/The Blain Model} & \colhead{Optical Thin/The Blain Model}
}

\startdata
GN10          & 53$\pm$2 & 40$\pm$2 & 67$\pm$3 & 41$\pm$2\\
GN20          & 46$\pm$1 & 36$\pm$1 & 54$\pm$2 & 36$\pm$1\\
GN20.2        & 86$\pm$5 & 67$\pm$4 & 71$\pm$5 & 57$\pm$3\\
AzTEC1        & 57$\pm$2 & 43$\pm$2 & 74$\pm$3 & 43$\pm$2\\
AzTEC3        & 79$\pm$7 & 54$\pm$5 & 103$\pm$9 & 24$\pm$1\\
Capak4.55     & 64$\pm$4 & 47$\pm$3 & 83$\pm$5 & 47$\pm$3\\
\enddata
\end{deluxetable}

\clearpage
\begin{deluxetable}{lcccccc}
\tabletypesize{\scriptsize}
\tablecaption{The Derived Parameters for this Sample\label{tbl-4}}
\tablewidth{0pt}
\tablehead{
\colhead{Name} & \colhead{Log($L_{\rm FIR}$/L$_{\odot}$)} & \colhead{T$_{dust}$ [K]} & \colhead{Log(L$_{4.5}$/L$_{\odot}$)} & \colhead{q} & \colhead{M$_{H_2}$(10$^{10}$M$_{\odot}$)} & \colhead{M$_{dust}$(10$^9$M$_{\odot}$)}
}
\startdata
GN10          & 12.85$\pm$0.03 & 53$\pm$2 & 11.13$\pm$0.08 & 2.27$\pm$0.12&2.7$\pm$0.5 &2.43$\pm$0.48\\
GN20          & 13.05$\pm$0.01 & 46$\pm$1 & 11.53$\pm$0.03 & 2.14$\pm$0.16&5.0$\pm$0.6 &5.24$\pm$0.81\\
GN20.2        & 12.83$\pm$0.03 & 86$\pm$5 & 11.02$\pm$0.07 & 1.55$\pm$0.05&3.0$\pm$1.0 &3.41$\pm$1.32\\
AzTEC1        & 13.17$\pm$0.03 & 57$\pm$2 & 11.52$\pm$0.05 & 2.25$\pm$0.29&~~          &3.65$\pm$0.71\\
AzTEC3        & 13.19$\pm$0.23 & 79$\pm$7 & $<$11.31       & $>$2.15      &5.3$\pm$0.6 &1.26$\pm$0.38\\
Capak4.55     & 12.95$\pm$0.26 & 64$\pm$4 & 11.25$\pm$0.22 & 2.14$\pm$0.17&2.6$\pm$0.5 &2.07$\pm$0.48\\
J033229  & 12.78$\pm$0.12 & ---- & 11.38$\pm$0.07 & 2.30$\pm$0.18&1.6$\pm$0.3&~~\\
\enddata

\end{deluxetable}


\begin{thebibliography}{}
\bibitem[Alexander et al.(2005)]{alexander2005}Alexander, D., et al. 2005, Nature, 434, 738
\bibitem[Allen et al.(2004)]{allen2004}Allen, L., et al. 2004, \apjs, 154, 363
\bibitem[Ashby et al.(2012)]{ashby2012}Ashby, M., et al. 2012, in preparation.
\bibitem[Baugh et al.(2005)]{baugh2005}Baugh, C. M., et al. 2005, \mnras, 356, 1191
\bibitem[Beelen et al.(2006)]{beelen2006}Beelen, A., et al. 2006, \apj, 642,694
\bibitem[Blain et al.(2002)]{blain2002} Blain, A. W., Smail, I., Ivison, R. J., Kneib, J.-P., \& Frayer, D. T.  2002, PhR, 369,111
\bibitem[Blain et al.(2003)]{blain2003} Blain, A. W., Barnard, V. E., Chapman, S. C.  2003, \mnras, 338, 733
\bibitem[Blain et al.(2004)]{blain2004} Blain, A. W., et al.  2004, \apj, 611, 725
\bibitem[Borys et al.(2005)]{borys2005}Borys, C., et al. 2005, \apj, 635,853
\bibitem[Bruzual \& Charlot (2003)]{bc03}Bruzual,  G. \& Charlot,  S. 2003,  \mnras,  344,  1000
\bibitem[Capak et al.(2008)]{capak2008}Capak, P. et al. 2008, \apj, 681, L53
\bibitem[Capak et al.(2011)]{capak2011}Capak, P. et al. 2011, Nature, 470,233
\bibitem[Carilli et al.(2010)]{2010ApJ...714.1407C} Carilli, C.~L., Daddi, E., Riechers, D., et al.\ 2010, \apj, 714, 1407 
\bibitem[Casey et al.(2012)]{casey2012}Casey, C. M., et al. 2012, \apj, 761, 139
\bibitem[Chakrabarti et al.(2008)]{chakrabarti2008}Chakrabarti, S., et al. 2008, \apj, 688, 972
\bibitem[Chapman et al.(2003)]{chapman2003}Chapman, S. C., Blain, A. W., Ivison, R. J., \& Smail, I. R. 2003, Nature, 422, 695
\bibitem[Clements et al.(2010)]{clements2010}Clements, D., Dunne, L., \& Eales, S. 2010, \mnras, 2010, 403,274
\bibitem[Condon et al.(1982)]{condon1982}Condon, J. J., et al. 1982, \apj, 252, 102
\bibitem[Conley et al.(2011)]{conley2011}Conley, A. et al. 2011, \apj, 732, 35
\bibitem[Coppin et al.(2008)]{coppin2008}Coppin, K. E. K., et al. 2008, \mnras, 389,45
\bibitem[Coppin et al.(2009)]{coppin2009}Coppin, K. E. K., et al. 2009, \mnras, 395,1905
\bibitem[Coppin et al.(2010)]{coppin2010}Coppin, K. E. K., et al. 2010, \mnras, 407, 103
\bibitem[Daddi et al.(2009a)]{daddi2009a}Daddi, E., Dannerbauer, H., Krips, M., Walter, F., Dickinson, M., Elbaz, D., \& Morrison, G. E. 2009a, \apj, 695, L176
\bibitem[Daddi et al.(2009b)]{daddi2009b}Daddi, E. et al. 2009b, \apj, 694, 1517
\bibitem[Daddi et al.(2010)]{daddi2010}Daddi, E. et al. 2010, \apjl, 714, L118
\bibitem[Dai et al.(2012)]{dai2011}Dai, Y., et al. 2012, \apj,753,33.
\bibitem[Dave et al.(2010)]{dave2010}Dave, R., et al. 2010, \mnras, 404,1355
\bibitem[Desai et al.(2009)]{desai2009}Desai, V., et al, 2009, \apj, 700,1190
\bibitem[Dey et al.(2008)]{dey2008}Dey, A., et al, 2008, \apj, 677, 943
\bibitem[Dowell et al.(2013)]{dowell2013}Dowell, C. D., et al. 2013, arXiv:1310.7583
\bibitem[Downes \& Solomon(1998)]{downes1998}Downes, D. \& Solomon, P. M. 1998, \apj, 507,615
\bibitem[Dye et al.(2008)]{dye2008}Dye, S., et al, 2008, \mnras, 386 1107 
\bibitem[Efstathiou et al.(1995)]{Efstathiou1995}Efstathiou A., Rowan-Robinson M., 1995, MNRAS 273, 649
\bibitem[Efstathiou et al.(2000)]{Efstathiou2000}Efstathiou A., Rowan-Robinson M., Siebenmorgen R., 2000, MNRAS 313, 734
\bibitem[Efstathiou et al.(2003)]{Efstathiou2003}Efstathiou A., Rowan-Robinson M., 2003, MNRAS 343, 322
\bibitem[Elvis et al.(1994)]{elvis1994}Elvis, M., et al. 1994, \apjs, 95, 1
\bibitem[Engelbracht et al.(2006)]{engelbracht2006}Engelbracht, C. W., et al. 2006, \apj, 642, L127
\bibitem[Farrah et al.(2008)]{farrah2008}Farrah, D., et al, 2008, \apj, 677,957 
\bibitem[F\"orster Schreiber et al.(2001)]{foster2001}F\"orster Schreiber, N. M., et al. 2001, \apj, 552, 544
\bibitem[Genzel et al.(2010)]{genzel2010}Genzel, R., et al. 2010, \mnras, 407,2091
\bibitem[Greve et al.(2005)]{greve2005}Greve, T. R., et al. 2005, \mnras, 359, 1165
\bibitem[Griffin et al. (2010)]{Griffin2010}Griffin, M.J., et al. 2010, \aap, 518, L3
\bibitem[Guo \& White (2008)]{guo2008}Guo, Q., \& White, S. D. M. 2008, \mnras, 384, 2
\bibitem[Haas et al.(2003)]{haas2003}Haas, M., et al. 2003, \aap, 402,87
\bibitem[Hatziminaoglou et al.(2010)]{hatzi2010}Hatziminaoglou, E., et al. 2010, \aap, 518, L33
\bibitem[Hodge et al.(2013)]{2013ApJ...776...22H} Hodge, J.~A., Carilli,C.~L., Walter, F., Daddi, E., \& Riechers, D.\ 2013, \apj, 776, 22 
\bibitem[Hopkins et al.(2009)]{2009ApJ...691.1168H} Hopkins, P.~F., Cox, T.~J., Younger, J.~D., \& Hernquist, L.\ 2009, \apj, 691, 1168
\bibitem[Huang et al.(2004)]{huang2004}Huang, J.-S., et al. 2004, \apjs, 154, 44
\bibitem[Huang et al.(2005)]{huang2005}Huang, J.-S., et al. 2005, \apj, 634,137
\bibitem[Huang et al.(2007)]{huang2007}Huang, J.-S., et al. 2007, \apj, 664,840
\bibitem[Huang et al.(2009)]{huang2009}Huang, J.-S., et al. 2009, \apj, 700,183
\bibitem[Huang et al.(2011)]{huang2011}Huang, J.-S., et al. 2011, \apj, 700,183
\bibitem[Houck et al.(2005)]{houck2005}Houck, J. et al. 2005, \apj, 622, L105
\bibitem[Iono et al.(2006)]{iono2006}Iono, D., Peck, A. B., Pope, A., Borys, C., Scott, D., Wilner, D. J., Gurwell, M., Ho, P. T. P., Yun, M. S., Matsushita, S., Petitpas, G. R., Dunlop, J. S., Elvis, M., Blain, A., \& Le Floc'h, E. 2006, \apj, 640,1
\bibitem[Ivison et al.(2011)]{2011MNRAS.412.1913I} Ivison, R.~J., Papadopoulos, P.~P., Smail, I., et al.\ 2011, \mnras, 412, 1913 
\bibitem[Klaas et al.(2001)]{klaas2001}Klaas, U., et al. 2001, \aap, 379, 823
\bibitem[Kovacs et al.(2006)]{kovacs2006}Kovacs, A., et al. 2006, \apj, 650,592
\bibitem[Kovacs et al.(2010)]{kovacs2010}Kovacs, A., et al. 2010, \apj, 717, 29
\bibitem[Laird et al.(2010)]{laird2010}Laird, E. S., et al. 2010, \mnras, 401, 2763
\bibitem[Lisenfeld et al.(2000)]{lisenfeld2000}Lisenfeld, U., et al. 2000, \mnras, 312, 433
\bibitem[Lutz et al.(2005)]{lutz2005}Lutz, D., et al. 2005, \apjl, 625, 83
\bibitem[Magdis et al.(2010)]{magdis2010}Magdis, G., et al. 2010, \mnras,409,22
\bibitem[Magdis et al.(2011)]{magdis2011}Magdis, G., et al. 2011, \apjl,740,L15
\bibitem[Magdis et al.(2012)]{magdis2012} Magdis, G.~E., Daddi, E., B{\'e}thermin, M., et al.\ 2012, \apj, 760, 6 
\bibitem[Magnelli et al.(2012)]{magnelli2012}Magnelli, B., et al. 2012,\aap,539,155
\bibitem[McCracken et al.(2010)]{mccracken2010}McCracken, H. J., et al. 2010,\apj,708, 202
\bibitem[Men�ndez-Delmestre et al.(2007)]{mend2007}Men�ndez-Delmestre, K., et al. 2007, \apj, 655L,65
\bibitem[Men�ndez-Delmestre et al.(2009)]{mend2009}Men�ndez-Delmestre, K., et al. 2009, \apj, 699,667
\bibitem[Narayanan et al.(2009)]{narayanan2009}Narayanan, D., et al. 2009, \mnras, 400, 1919
\bibitem[Narayanan et al.(2010)]{narayanan2010}Narayanan, D., et al. 2010, \mnras, 401, 1613
\bibitem[Nguyen et al.(2010)]{nguyen2010}Nguyen, H.T., et al. 2010, \aap,518, 5
\bibitem[Oliver et al.(2010)]{oliver2010}Oliver, S., et al. 2010, \aap,518,21
\bibitem[Oliver et al.(2012)]{2012MNRAS.424.1614O} Oliver, S.~J., Bock, J., Altieri, B., et al.\ 2012, \mnras, 424, 1614
\bibitem[Pilbratt et al.(2010)]{pilbratt2010}Pilbratt, G. L., et al. 2010, \aap, 518,1
\bibitem[Pope et al.(2008)]{pope2008}Pope, A., et al. 2008, \apj, 675,1171
\bibitem[Riechers et al.(2010)]{riechers2010}Riechers, D. A. et al. 2010, \apj, 720, L131
\bibitem[Riechers et al.(2011)]{2011ApJ...739L..32R} Riechers, D.~A., Carilli, C.~L., Maddalena, R.~J., et al.\ 2011, \apjl, 739, L32 
\bibitem[Riechers et al.(2013a)]{2013arXiv1306.5235R} Riechers, D.~A., Pope, A., Daddi, E., et al.\ 2013, arXiv:1306.5235 
\bibitem[Riechers et al.(2013b)]{Riechers2013b}Riechers, D.~A., Bradford, C. M., Clements, D. L., et al.\ 2013, Nature, 496 329
\bibitem[Rigby et al.(2008)]{rigby2008}Rigby, J., et al. 2008, \apj, 675,262
\bibitem[Rigopoulou et al.(2010)]{rigopoulou2010}Rigopoulou, D., et al. 2010,\mnras, 409, 7
\bibitem[Roseboom et al.(2010)]{roseboom2010}Roseboom, I., et al. 2010, \mnras, 409, 48
\bibitem[Roseboom et al.(2012)]{roseboom2012}Roseboom, I., et al. 2012, \mnras, 419,2758
\bibitem[Rowan-Robinson(1995)]{Rowan-Robinson1995}Rowan-Robinson M., 1995, MNRAS 272, 737
\bibitem[Rowan-Robinson et al.(2008)]{Rowan-Robinson2008}Rowan-Robinson M. et al, 2008, MNRAS 386, 697
\bibitem[Sanders et al.(2007)]{sanders2007}Sanders, D., et al. 2007, \apjs, 172,86
\bibitem[Schinnerer et al.(2008)]{schinnerer2008}Schinnerer, E., et al. 2008, \apj, 689, L5
\bibitem[Shu et al.(2001)]{shu2001}Shu, C., et al., 2001, \mnras, 327,895
\bibitem[Smith et al.(2010)]{smith2010}Smith, A., et al., 2010, \mnras in press
\bibitem[Smolcic et al.(2010)]{smolcic2011}Smolcic, V., et al. 2011, \apj, 731, L27
\bibitem[Solomon \& Vanden Bout(2005)]{solomon2005}Solomon, P. M. \& Vanden Bout, P. A. 2005, \araa, 43,677
\bibitem[Swinbank et al.(2006)]{swinbank2006}Swinbank, A. M., et al. 2006, \mnras, 371, 465
\bibitem[Swinbank et al.(2008)]{swinbank2008}Swinbank, A. M., et al. 2008, \mnras, 391, 420
\bibitem[Teplitz et al.(2006)]{teplitz2006}Teplitz, H., et al. 2006, in BAAS, 38, 1079
\bibitem[valiante et al.(2007)]{valiante2007}Valiante, E., Lutz, D., Sturm, E., Genzel, R., Tacconi, L. J., Lehnert, M. D., \& Baker, A. J. 2007, \apj, 660, 1060
\bibitem[Wang et al.(2007)]{wang2007}Wang, W.-H., Cowie, L. L., van Saders, J., Barger, A. J., \& Williams, J. P. 2007, \apj, 670, 89
\bibitem[Wang et al.(2009)]{wang2009}Wang, W.-H., Barger, A. J., \& Cowie, L. L. 2009, \apj, 690,319
\bibitem[Wang et al.(2011)]{wangren2011}Wang, R., et al. 2011, \aj, 142, 101
\bibitem[Yan et al.(2007)]{yan2007}Yan, L., et al. 2007, \apj, 658, 778
\bibitem[Yan et al.(2010)]{yan2010}Yan, L., et al. 2010, \apj, 714, 100
\bibitem[Yang et al.(2007)]{yang2007}Yang,  M., Greve, T. R., Dowell, C. D., \& Borys, C. 2007, \apj, 2007, 660,1198
\bibitem[Younger et al.(2007a)]{younger2006}Younger. J. D., et al. 2007a, \apj, 671, 1241
\bibitem[Younger et al.(2007b)]{younger2007}Younger. J. D., et al. 2007b, \apj, 671, 1531




\end{thebibliography}
\end{document}